\documentclass[pra,twocolumn,showpacs,preprintnumbers,tightenlines,epsfig,floats,superscriptaddress]{revtex4}
\usepackage{graphicx}
\usepackage{dcolumn}
\usepackage{bm}
\usepackage{amsmath}
\usepackage{amsfonts}
\usepackage{amssymb}
\usepackage{times}

\begin{document}

\title{Pairing mean-field theory for the dynamics of dissociation of molecular Bose-Einstein condensates}
\author{M. J. Davis}
\affiliation{ARC Centre of Excellence for Quantum-Atom Optics, School of Physical
Sciences, University of Queensland, Brisbane, Qld 4072, Australia}
\author{S. J. Thwaite}
\affiliation{ARC Centre of Excellence for Quantum-Atom Optics, School of Physical
Sciences, University of Queensland, Brisbane, Qld 4072, Australia}
\affiliation{Department of Physics, University of Auckland, Private Bag 92019, Auckland,
New Zealand}
\author{M. K. Olsen}
\affiliation{ARC Centre of Excellence for Quantum-Atom Optics, School of Physical
Sciences, University of Queensland, Brisbane, Qld 4072, Australia}
\author{K. V. Kheruntsyan}
\affiliation{ARC Centre of Excellence for Quantum-Atom Optics, School of Physical
Sciences, University of Queensland, Brisbane, Qld 4072, Australia}
\date{\today {}}

\begin{abstract}
We develop a pairing mean-field theory to describe the quantum
dynamics of the dissociation of molecular Bose-Einstein condensates
into their constituent bosonic or fermionic atoms. We apply the
theory to one, two, and three-dimensional geometries and analyze the
role of dimensionality on the atom production rate as a function of
the dissociation energy. As well as determining the populations and
coherences of the atoms, we calculate the correlations that exist
between atoms of opposite momenta, including the column density
correlations in 3D systems. We compare the results with those of the
undepleted molecular field approximation and argue that the latter
is most reliable in fermionic systems and in lower dimensions. In
the bosonic case we compare the pairing mean-field results with
exact calculations using the positive-$P$ stochastic method and
estimate the range of validity of the pairing mean-field theory.
Comparisons with similar first-principle simulations in the
fermionic case are currently not available, however, we argue that
the range of validity of the present approach should be broader for
fermions than for bosons in the regime where Pauli blocking prevents
complete depletion of the molecular condensate.
\end{abstract}


\maketitle

\section{Introduction}

The generation and detection of strongly correlated atomic ensembles
is becoming one of the central themes in the study of ultracold quantum
gases, leading to the birth of the subfield of quantum-atom optics. Recently
there have been an increasing number of experiments probing the higher-order
coherences and correlation functions that are fundamental to this field and are
analogous to similar developments in the early days of quantum optics with
photons.

One of the first experiments in this area was the measurement of a
two-time second-order correlation function in a thermal gas of
metastable Neon atoms falling on four microchannel plate (MCP)
detectors \cite{Yasuda-Shimizu}. Another early experiment was the
measurement of the rate of three-body losses from a Bose-Einstein
condensate (BEC), from which the local third-order correlation
function could be inferred \cite{burt97} (see also \cite{tolra04}).
More recently, the arsenal of experimental techniques to study
higher-order correlations has been expanded to include the
measurement of photoassociation rates \cite{kinoshita05}, shot-noise
spectroscopy of absorption images
\cite{Greiner,Bloch,Bloch-HBT-fermions,esteve06}, atom counting
using fluorescence imaging and high-finesse optical cavities
\cite{Raizen,Esslinger}, and position-resolved counting of
metastable Helium
atoms using arrays of MCP detectors \cite%
{Aspect,Aspect-HBT-fermions,Perrin-BEC-collisions}.
Proposals for additional techniques include
 atom counting using photoionization \cite%
{photoionization}, and freeze-in techniques with optical lattices
for \textit{in-situ} measurements of non-local spatial correlations
\cite{Sykes}.

The experimental systems that are being studied with these new
techniques for correlation measurements include Bose-Einstein
condensates and degenerate Fermi gases
\cite{burt97,Aspect,Aspect-HBT-fermions}, atom lasers
\cite{Esslinger}, one-dimensional (1D) and quasi-1D Bose gases
\cite{tolra04,kinoshita05,esteve06,Raizen}, Bose and Fermi gases in
optical lattices \cite{Bloch,Bloch-HBT-fermions}, four-wave mixing
via BEC collisions \cite{Perrin-BEC-collisions} and molecular
dissociation of BECs \cite{Greiner}. The last two systems are
relevant to the present work and have similar atom-atom correlations
on the dissociation sphere and on the $s$-wave scattering halo,
respectively \cite{Trippenbach,Perrin-theory}. All the experiments
described above generate quantum correlations that cannot be
described by Gross-Pitaevskii mean-field theory.    They provide a
challenge for theorists to develop and apply more advanced
techniques in order to quantitatively model experimental systems,
such as used in
Refs.~\cite{Moelmer2001,twinbeams,Trippenbach,Vardi-Moore,Norrie-Ballagh-Gardiner,
Rey-Clark,Jack-Pu,Savage,DeuarDrummond-4WM,Challis}.

Particular interest in molecular dissociation
experiments  stems from the prospect of
generating and detecting squeezed states of matter waves \cite%
{Kasevich,Moelmer2001,Roberts,twinbeams,Yurovski-diss,Haine,SavageKheruntsyanSpatial},
with potential applications in precision measurements.  There also exists
the possibility of  producing Einstein-Podolsky-Rosen
correlations and other
entangled states of massive particles \cite%
{Meystre-spin-EPR,Duan-spin-EPR,Soerensen-Duan-Zoller,Yurovski-FWM,KPRL,Hope,Olsen-Davis,Zhao-Astrakharchik}
for fundamental tests of quantum theory in mesoscopic regimes and
for quantum information science applications.
Additionally, molecular dissociation can serve as a probe of two-body
interactions, collisional resonances, and spectroscopic properties
of Feshbach resonance molecules \cite%
{Wieman-Julienne-dissociation,Rempe-Kokkelmans-dissociation,Braaten-2006,Hanna-2006}.
Other recent studies of dissociation are concerned with the role of
confinement on the stability of a molecular BEC against dissociation \cite%
{Vardi}, the effect of magnetic field fluctuations on dissociation \cite%
{Plata-2005}, and the dynamics of molecule-atom conversion in
optical lattices \cite{Meystre-diss}.

The purpose of this paper is to develop and apply a pairing
mean-field theory (PMFT) to the dynamics of dissociation of a BEC of
molecular dimers consisting of pairs of either bosonic or fermionic
atoms. The mean fields for the molecules are introduced in the
standard way at the level of amplitudes, whereas for the atoms they
are introduced at the level of pairing fields \cite{Holland} that
describe normal and anomalous moments of the atomic creation and
annihilation operators. This approach takes into account the
molecular depletion and can facilitate the calculation of atom-atom
pair correlation functions beyond the coherent level of the simple
mean field theory. The present PMFT is similar to that of Jack and
Pu \cite{Jack-Pu}, except that: (\textit{i}) we extend the analysis
to one-dimensional (1D) and two-dimensional (2D) systems;
(\textit{ii}) we relax the approximation of the constant density of
atomic states around the peak dissociation energy; and
(\textit{iii}) we calculate the pair correlation functions for
dissociated atoms.

First-principle quantum simulations of the dissociation of molecular
BECs into bosonic atoms have been performed using the stochastic
positive-$P$ representation method
\cite{Moelmer2001,twinbeams,Savage}, however this technique has some
limitations. Most significant is the short time scale over which the
positive-$P$ method remains valid and produces results free of large
sampling errors or the boundary term problem \cite{Gilchrist}. For
pure dissociation, without $s$-wave scattering interactions, the
typical timescale for the positive-$P$ simulations is limited to
about $50\%$ conversion \cite{Savage}. The addition of $s$-wave
scattering interactions further limits this to even shorter
durations and only $5\%-10\%$ conversion for typical experimental
parameters \cite{Savage}. The situation is less clear in the
fermionic case; even though similar first-principle techniques for
simulating multi-mode fermion dynamics have recently been developed
\cite{Corney-fermionic}, these techniques have not been
applied yet to the problem of multi-mode dissociation for
quantitative comparison.

Given the limitations of exact techniques, it is useful to develop
alternative, approximate theoretical techniques that can deal with
longer durations of dissociation and larger conversion percentage.
The PMFT developed here and in Ref.~\cite{Jack-Pu} falls into this
category and goes beyond previous approximate theories that ignore
molecular depletion altogether \cite{Savage,Fermidiss}. As such, the
range of its validity with respect to molecular depletion covers
dissociation durations corresponding to near complete conversion. In
addition, the PMFT is able to treat the dissociation into bosonic
and fermionic atoms on equal footing and therefore is closer to
being
able to quantitatively describe the experiments on dissociation of $%
^{40}$K$_{2}$ dimers into fermionic atoms \cite{Greiner}, in
addition to the earlier experiments on dissociation of
$^{23}$Na$_{2}$ \cite{Dissociation-exp-Ketterle} and $^{87}$Rb$_{2}$
\cite{Durr} dimers into bosonic atoms.

This paper is organized as follows. In Sec.~\ref{sect:model} we
introduce the model Hamiltonians describing the dissociation into
distinguishable and indistinguishable atoms and discuss the validity
of approximations involved in our model. In Sec.~\ref{sect:DecayRate}
we give the simplest possible description of
dissociation based on a Fermi's golden rule calculation of the
molecular decay rate and discuss the implications for dissociation
in 1D, 2D and 3D geometries. In Sec. \ref{sect:PMFT} we develop the
PMFT and discuss our main results for dissociation dynamics, atomic
momentum distribution, and atom-atom correlations. Our conclusions
are summarized in Sec. \ref{sect:Conclusions}.

\section{The model}

\label{sect:model}

In our treatment of dissociation we assume an initial condition of
a stationary BEC of diatomic molecules consisting of either
(\textit{i}) two distinguishable fermionic or bosonic atoms (e.g.
in two different spin states) or (\textit{ii}) two indistinguishable
bosonic atoms. In the case (\textit{i}), we further assume that both
constituents are of the same species and the same mass; we do not consider
heteronuclear molecules. The model Hamiltonians for each of
these situations are described below, and we discuss the initial
conditions and the validity of our approximations in Sec.
\ref{sect:approximations}.

\subsection{Dissociation into distinguishable atoms}

The quantum field theoretic effective Hamiltonian describing the system in one,
two, or three ($D=1,2,3$) spatial dimensions is given, in a rotating frame, by
\cite{JOptB1999-PRA2000}%
\begin{align}
\hat{H}& =\int d^{D}\mathbf{x}\left\{ \sum\limits_{i=0,1,2}\frac{\hbar ^{2}}{%
2m_{i}}|\mathbf{\nabla }\hat{\Psi}_{i}|^{2}+\hbar \Delta (\hat{\Psi}%
_{1}^{\dagger }\hat{\Psi}_{1}+\hat{\Psi}_{2}^{\dagger }\hat{\Psi}_{2})\right.
\notag \\
& \left. -i\hbar \chi _{D}\left( \hat{\Psi}_{0}^{\dagger }\hat{\Psi}_{1}\hat{%
\Psi}_{2}-\hat{\Psi}_{2}^{\dagger }\hat{\Psi}_{1}^{\dagger }\hat{\Psi}%
_{0}\right) \right\} .  \label{eq:Ham}
\end{align}%
The molecular field is described by a bosonic operator $\hat{\Psi}_{0}(%
\mathbf{x},t)$ whilst the two atomic fields are described by either
bosonic \emph{or} fermionic operators $\hat{\Psi}_{1}(\mathbf{x},t)$ and $%
\hat{\Psi}_{2}^{\dagger }(\mathbf{x},t)$. The field operators satisfy the
respective commutation or anti-commutation relations, i.e., $[\hat{\Psi}_{i}(%
\mathbf{x},t),\hat{\Psi}_{j}^{\dagger }(\mathbf{x}^{\prime },t)]=\delta
_{ij}\delta ^{(D)}(\mathbf{x}-\mathbf{x}^{\prime })$ in the case of bosonic
atoms or  $\{\hat{\Psi}_{i}(\mathbf{x},t),\hat{\Psi}_{j}^{\dagger }(%
\mathbf{x}^{\prime },t)\}=\delta _{ij}\delta ^{(D)}(\mathbf{x}-\mathbf{x}%
^{\prime })$ ($i=1,2$) in the fermionic case. For notational simplicity, we
use $\mathbf{x}$ for the position in 1D, 2D and 3D cases, with the
understanding that in the 1D case it is a scalar.

The first term in the Hamiltonian (\ref{eq:Ham}) describes the
kinetic energy where the atomic masses are $m_{1}=m_{2}$, whereas
the molecular mass is $m_{0}=m_{1}+m_{2}=2m_{1}$. The coupling term
$\chi _{D}$ is responsible for coherent conversion of molecules into
atom pairs, e.g. via optical
Raman transitions or a Feshbach resonance sweep (see, for example, Refs. \cite%
{JOptB1999-PRA2000,PDKKHH-1998,Superchemistry,Feshbach-KKPD,Timmermans,JJ-1999,Holland}
and \cite{Stoof-review,Julienne-review} for recent reviews). For the
Raman case, the coupling $\chi_{3D}$ is expressed in terms of the
Rabi frequencies for free-bound and bound-bound transitions as in
Ref.~\cite{Superchemistry}.

In the case of a Feshbach resonance, the coupling $\chi _{3D}$ is
given by (see \cite{Holland,Timmermans,Feshbach-KKPD} for
notational consistency)%
\begin{equation}
\chi _{3D}=\sqrt{4\pi a_{bg}\Delta \mu \Delta B/m_{1}}.  \label{chi-3D}
\end{equation}%
Here $\Delta \mu =\mu _{1}+\mu _{2}-\mu _{0}$ is the difference in
the magnetic moments of the atomic and the bound molecular states,
$\Delta B$ is the magnetic width of the resonance, and $a_{bg}$ is
the background scattering length for $s$-wave collisions of the
atoms in the two spin states. In systems of reduced dimensionality
(1D or 2D) and away from confinement induced resonances, the
couplings $\chi _{1D}$ and $\chi _{2D}$, are obtained by integrating
over the ground-state wave function in the tightly confined
dimensions. Assuming harmonic trapping potentials in the eliminated
dimensions, with oscillation frequencies $\omega _{\perp }\equiv
\omega _{y}=\omega _{z}$ in the 1D case and $\omega _{z}$ in the 2D
case,
the respective atom-molecule couplings are%
\begin{eqnarray}
\chi _{1D} &=&\chi _{3D}/\left( 2\pi l_{\perp }^{2}\right) ^{1/2},
\label{chi-1D} \\
\chi _{2D} &=&\chi _{3D}/\left( 2\pi l_{z}^{2}\right) ^{1/4}.  \label{chi-2D}
\end{eqnarray}%
Here $l_{\perp }=\sqrt{\hbar /m_{1}\omega _{\perp }}$ and $l_{z}=\sqrt{\hbar
/m_{1}\omega _{z}}$ are the harmonic oscillator lengths for the atoms. For a
given transverse trap  frequency $\omega _{\perp }$ in the 1D
geometry, the dissociation detuning must satisfy $|\Delta |<\omega _{\perp }$
to avoid transverse excitations and maintain the validity of the 1D
treatment. Similarly, in the 2D case one has to satisfy $|\Delta |<\omega
_{z}$ in order that the system remains in the 2D regime in the $xy$ plane.

The quantity $\Delta$ in the Hamiltonian (\ref{eq:Ham}) is the
effective detuning between the molecular state with energy $E_0$ and
the atomic states with energies $E_1$ and $E_2$. It is defined
differently depending on whether the coupling is due to a sweep
through a Feshbach resonance, or due to Raman photoassociation
lasers.

\begin{itemize}
\item[(\textit{i})] In the Feshbach resonance case, dissociation
of an initially stable ($E_{0}<E_{1}+E_{2}$) molecular BEC into the
constituent atoms may be achieved by a rapid magnetic field sweep
onto the atomic side of the resonance ($E_{0}>E_{1}+E_{2}$)
\cite{Durr}. The detuning is defined as the overall energy mismatch
$2\hbar \Delta =E_{1}+E_{2}-E_{0}$ between the free two-atom state
in the dissociation threshold with energy $E_{1}+E_{2}$ and the
bound molecular state of energy $E_{0}$. The sweep is implemented to
result in a finite and negative detuning $2\hbar \Delta <0$ after the sweep (%
$E_{0}>E_{1}+E_{2}$), in which case the molecules become unstable against
spontaneous dissociation into free atom pairs.

\item[(\textit{ii})] In the case of two-photon Raman photoassociation \cite%
{twinbeams,Superchemistry}, the overall energy mismatch is given by $2\hbar
\Delta =E_{1}+E_{2}-E_{0}-\hbar \omega $, where $\omega $ is the frequency
difference between the two Raman lasers. Similarly, for an rf transition
\cite{Greiner} $\omega $ is the rf frequency.
\end{itemize}

The Hamiltonian (\ref{eq:Ham}) conserves the total number of atomic particles%
\begin{equation}
\hat{N}=2\hat{N}_{0}(t)+\hat{N}_{1}(t) +\hat{N}_{2}(t)=\mathrm{const},
\label{eq:Ncons_dist}
\end{equation}%
where the constant is given by $2\hat{N}_{0}(0)$ for the vacuum
initial condition for the atoms.

\subsection{Dissociation into indistinguishable atoms}

Molecular dissociation into pairs of indistinguishable bosonic atoms in the
same internal state is described by the following Hamiltonian \cite%
{PDKKHH-1998}, in a rotating frame
\begin{align}
\hat{H}& =\int d^{D}\mathbf{x}\left\{ \sum\limits_{i=0,1}\frac{\hbar ^{2}}{%
2m_{i}}|\mathbf{\nabla }\hat{\Psi}_{i}|^{2}+2\hbar \Delta \hat{\Psi}%
_{1}^{\dagger }\hat{\Psi}_{1}\right.  \notag \\
& \left. -i\frac{\hbar \chi _{D}}{2}\left( \hat{\Psi}_{0}^{\dagger }\hat{\Psi%
}_{1}^{2}-\hat{\Psi}_{1}^{\dagger 2}\hat{\Psi}_{0}\right) \right\} .
\label{Hindistinguish}
\end{align}%
where $\hat{\Psi}_{1}(\mathbf{x})$ is the atomic field operator, and $\chi
_{D}$ is the atom-molecule coupling. The microscopic expression for the 3D
coupling $\chi _{3D}$ in the case of a Feshbach resonance is given by \cite%
{Holland,Timmermans,Feshbach-KKPD}%
\begin{equation}
\chi _{3D}=\sqrt{8\pi a_{bg}\Delta \mu \Delta B/m_{1}},
\label{chi3D-homonuclear}
\end{equation}%
where $a_{bg}$ is the atom-atom background scattering length,
$\Delta \mu =2\mu _{1}-\mu _{0}$ is the magnetic moment difference
between the atomic and the bound molecular states, and $\Delta B$ is
the magnetic width of the resonance. For systems of reduced
dimensionality, Eqs.~(\ref{chi-1D}) and (\ref{chi-2D}) for
$\chi _{1D}$ and $\chi _{2D}$ are unchanged.

The Hamiltonian (\ref{Hindistinguish}) conserves the total number of atomic
particles%
\begin{equation}
\hat{N}=2\hat{N}_{0}(t)+\hat{N}_{1}(t)=\mathrm{const},
\label{eq:Ncons_indist}
\end{equation}%
where the constant is given by $2\hat{N}_{0}(0)$ for the vacuum
initial condition for the atoms.

\subsection{Initial conditions and approximations}

\label{sect:approximations}

Starting with a stable molecular condensate, we assume that the coupling $%
\chi _{D}$ is switched on in the regime of a sudden jump \cite{Hanna-2006};
initially the molecules are assumed to be in a coherent state, whereas the
atomic fields are in the vacuum state. The energy level configuration \emph{after}
the Feshbach sweep or the Raman transitions (corresponding to $\Delta <0)$
is the actual initial condition for our simulations.

For a molecule at rest, the excess of potential energy $2\hbar |\Delta |$
(which we also refer to as the dissociation energy) is converted into
kinetic energy $2\hbar ^{2}k^{2}/(2m_{1})$ of atom pairs with equal but
opposite momenta $\pm \mathbf{k}_{0}$, where $k_{0}=|\mathbf{k}_{0}|=\sqrt{%
2m_{1}|\Delta |/\hbar }$. This is the physical origin of the expected
correlations between the atoms, which we will study below in the context of
many-body field theory.

The main limitation of our treatment is that we consider a spatially
uniform system in a cubic box with periodic boundary conditions.
The question that one has to address then is how well the results of
a uniform model can describe realistic nonuniform systems. It has previously
been shown  \cite{Savage} that while the uniform treatment can
give quantitatively reasonable results for the total number of atoms
and the atomic density distribution (provided the uniform system is
appropriately size-matched to a nonuniform system, as in Sec.
\ref{sect:total-N}), such a treatment is not adequate for
giving correct quantitative results for density-density correlation
functions. The reason is the mode-mixing due to inhomogeneity of
trapped condensates, which can strongly degrade the correlations
compared to the predictions of the uniform model \cite{Savage}.
Nevertheless, the merit of calculating the correlation functions
here is that the results provide  upper bounds for the correlations \cite%
{Fermidiss,Savage} and hence aid qualitative understanding.

Another obvious consideration when applying the uniform results
to size-matched nonuniform systems is that the duration of
dissociation should not exceed the time required for the atoms to
propagate distances larger than the size of the uniform box $L$.  Thus
the results are valid for times
\begin{equation}
t\lesssim t_{\max }=\frac{L}{v_{0}}=L\sqrt{m_{1}/(2\hbar |\Delta |)},
\label{box-time}
\end{equation}%
where $v_{0}=\hbar k_{0}/m_{1}$ is the mean velocity of the
dissociated atoms. In other words, while our numerical results can
be formally obtained for arbitrarily long times, the above equation
should be used to cutoff these results at $t_{\max }$ when they are
applied to a specific nonuniform system of size $L$.

The next major assumption in our treatment is the mean-field coherent state
for molecules at all times \cite{Jack-Pu}. This approximation breaks down
once the dissociation approaches the regime of complete depletion of the
molecular condensate ($\sim100\%$ conversion) when quantum fluctuations
become increasingly important. Accordingly, the results of the mean field
theory can not be trusted past that point in time; we present these results
only for academic purposes. We note that nearly complete conversion in the
mean field theory would always take place in the case of bosonic atoms, but
not necessarily in the fermionic case where the Pauli exclusion principle for
the atoms can dominate the molecular depletion. In this situation the
fermionic results can prove to be reliable for longer durations than in the
bosonic case.

Finally, we do not include any $s$-wave scattering interaction
terms. Potentially these interactions can affect the atom-atom
correlations in realistic nonuniform systems, especially in the long
time limit. However, their effect is negligible at low particle
densities and short dissociation times \cite{Savage}. At the level
of mean fields, the $s$-wave scattering terms can be neglected if
the dissociation energy is much larger than the mean-field
interaction energy. More importantly, it has been shown in Ref.
\cite{Savage} that the strongest degradation of atom-atom
correlations in realistic systems (compared to the predictions of
uniform models) comes not from the $s$--wave scattering
interactions, but from the mode-mixing due to inhomogeneity.
Accordingly, any theory that attempts to give quantitatively correct
results for atom-atom correlations has to first address the question
of inhomogeneity before including $s$-wave
interactions. The pairing mean field theory developed here does not
have this goal, but only represents an intermediate step between the
simple analytic theory of undepleted molecules
\cite{Fermidiss,Savage} and a more complete theory that can reliably
treat mode-mixing due to inhomogeneity and $s$-wave scattering
interactions for both bosonic and fermionic atoms.

\section{Decay rate and dissociation dynamics from Fermi's golden rule}

\label{sect:DecayRate}

\subsection{Distinguishable atoms}

Before presenting our numerical results within the framework of the
PMFT, we discuss the simpler results that follow from a Fermi's golden
rule calculation of the molecular decay rate. Due to the
conservation of the total number of atomic particles, the molecular
decay rate can be converted into the rate of atom production and
serve as the simplest description of the dynamics of dissociation in
the initial spontaneous regime. In particular, Fermi's golden rule
gives the correct initial behavior of the total atom number as a
function of time since at this stage neither bosonic stimulation nor
Pauli blocking affects the dynamics of individual modes. It also
 provides simple and useful insights
into the differences associated with the dimensionality of the
system, which are not explicit in the numerical results.
The derivation presented here is a direct calculation
based on the actual many-body Hamiltonian rather than an estimate
based on the comparison of the mean-field energy shift and the
energy shift of a Feshbach resonance state as in Ref.~\cite{Dissociation-exp-Ketterle}.

According to Fermi's golden rule, the molecular decay rate $\Gamma $ is
given by \cite{Dissociation-exp-Ketterle,Mies}%
\begin{equation}
\Gamma =\frac{2\pi }{\hbar }|V_{ma}|^{2}D^{(2)}(\epsilon ),
\label{Gamma-def}
\end{equation}%
where $D^{(2)}(\epsilon )$ is the density of two-atom states, $\epsilon $ is
the total kinetic energy of two atoms corresponding to the total
dissociation energy $\epsilon =2\hbar |\Delta |$, and $V_{ma}$ is the
transition matrix element for the atom-molecule coupling channel.
Calculating $V_{ma}$ explicitly from the interaction Hamiltonian in Eq.~(%
\ref{eq:Ham}) (see Appendix \ref{sect:appendix-A} for details), we obtain
the following results in 1, 2, and 3 dimensions:%
\begin{equation}
\Gamma =\left\{
\begin{array}{ll}
\lambda \chi_{1D}^{2}/ |\Delta|^{1/2},&\mathrm{(1D)}, \\
\\
\lambda^2 \chi_{2D}^{2},&\mathrm{(2D)},
\\\\
2 \lambda^3 \chi_{3D}^{2}|\Delta |^{1/2}/\pi,& \mathrm{(3D)}%
\end{array}%
\right.  \label{Gamma}
\end{equation}
where $\lambda = (m_1/2\hbar)^{1/2}$ is a constant. Since each
molecule produces two atoms (one per spin state) we can treat the
decay rate $\Gamma $ as the atom production rate using the conserved
total particle number
$2N_{0}(t)+N_{1}(t)+N_{2}(t)=2N_{0}(0)=\mathrm{const}$. Here,
$N_{i}=\langle \hat{N}_{i} \rangle$ with $N_{0}(0)$ being the total
initial number of molecules with no atoms present. Thus, the
solution $N_{0}(t)=N_{0}(0)\exp (-\Gamma t)$ to the
rate equation for the molecular decay, can be rewritten as $%
N_{1}(t)+N_{2}(t)=2N_{0}(0)\left[ 1-\exp (-\Gamma t)\right] $.
Noting also that $N_{1}(t)=N_{2}(t)$ since $\hat{N}_{1}-\hat{N}_{2}$
is conserved, the final result for the total number of atoms in each
spin state takes the
following simple form:%
\begin{equation}
N_{1(2)}(t)=N_{0}(0)( 1-e^{-\Gamma t})\simeq N_{0}(0)\Gamma t,
\;\;(\Gamma t\ll 1). \label{Atom-number}
\end{equation}%
We see immediately that the atom production rate in 1D, 2D and 3D
increases, remains unchanged and decreases, respectively, with the
dissociation energy $2\hbar |\Delta|$, following the respective
dependence of the decay rate $\Gamma$ on the detuning $|\Delta|$.
This is a direct consequence of the different dependence of the
density of atomic states $D(E)$ on energy $E$ in 1D, 2D and 3D.

\subsection{Indistinguishable atoms}

The decay rate $\Gamma $ derived above is for dissociation into
\textit{distinguishable} atoms. For completeness, we now present the
derivation in the case of \textit{indistinguishable} (bosonic) atoms
and show that the resulting expression in 3D coincides with the
expression given in Ref.~\cite{Dissociation-exp-Ketterle}.
The reasons for presenting these details are two-fold. First, we would
like to point out that even though the present paper is primarily
concerned with dissociation into distinguishable atom pairs (either
fermions or bosons in different spin states), the results in the
bosonic case can be easily modified to describe the case of
indistinguishable atoms in the same spin state. Secondly, we would
like to spell out the connection of the microscopic Feshbach
parameters with the coupling constant $\chi _{3D}$ in the respective
field-theory Hamiltonians without leaving room for confusion about
\textquotedblleft factors of two\textquotedblright . The details of
derivation of the decay rate $\Gamma ^{\prime }$ in the present case
are
given in Appendix \ref{sect:appendix-B}; the final result is:%
\begin{equation}
\Gamma^{\prime } =\left\{
\begin{array}{ll}
\lambda \chi_{1D}^{2}/ 2|\Delta|^{1/2},&\mathrm{(1D)}, \\
\\
\lambda^2 \chi_{2D}^{2}/2,&\mathrm{(2D)},
\\\\
\lambda^3 \chi_{3D}^{2}|\Delta |^{1/2}/\pi,& \mathrm{(3D)}%
\end{array}%
\right.  \label{Gamma'}
\end{equation}
For the same coupling strength $\chi _{D}$ these rates appear to be
simply half the corresponding result for the distinguishable case, Eq.~(%
\ref{Gamma}). However, one has to take into account that the
interaction Hamiltonian for the present indistinguishable case has
an extra factor of $1/2$ in front of $\chi _{D}$ and that the
definition of $\chi _{D}$ in the two cases are different.

The total number of atoms produced during the initial stage of
dissociation ($\Gamma ^{\prime }t\ll 1$) is
\begin{equation}
N_{1}(t)= 2N_{0}(0)(1-e^{-\Gamma ^{\prime }t})\simeq
2N_{0}(0)\Gamma^{\prime }t, \;(\Gamma^{\prime } t\ll 1). \label{N3D}
\end{equation}%
Comparing this with the result of Eq.~(\ref{Atom-number}) we see that for
the same coupling constant $\chi _{D}$ in the Hamiltonians~(\ref%
{Hindistinguish}) and (\ref{eq:Ham}), the total number of atoms produced in
the indistinguishable case corresponds to the number of atoms in \emph{each}
spin state produced in the distinguishable case.

\section{Pairing mean-field theory}

\label{sect:PMFT}

\subsection{Distinguishable atoms}

Considering a uniform system in a cubic box of volume $V=$ $L^{3}$ (or a
square of area $A=L^{2}$ in 2D, or a line of length $L$ in 1D), we expand
the molecular and atomic field operators in terms of plane-wave modes%
\begin{eqnarray}
\hat{\Psi}_{0}(\mathbf{x},t) &=&\frac{1}{L^{D/2}}\sum\nolimits_{\mathbf{k}}
\hat{b}_{\mathbf{k}}(t)e^{i\mathbf{k\cdot x}}, \\
\hat{\Psi}_{j}(\mathbf{x},t) &=&\frac{1}{L^{D/2}}\sum\nolimits_{\mathbf{k}}
\hat{a}_{j,\mathbf{k}}(t)e^{i\mathbf{k\cdot x}},
\end{eqnarray}%
where $\hat{b}_{\mathbf{k}}(t)$ and $\hat{a}_{j,\mathbf{k}}(t)$ ($j=1,2$)
are the corresponding creation operators satisfying the usual bosonic
commutation relation for molecules $[\hat{b}_{\mathbf{k}},\hat{b}_{\mathbf{k}
^{\prime }}^{\dagger }]=\delta _{\mathbf{k},\mathbf{k}^{\prime }}$ and
fermionic anti-commutation or bosonic commutation relations for the atoms, $%
\{\hat{a}_{i,\mathbf{k}},\hat{a}_{j,\mathbf{k}^{\prime }}^{\dagger
}\}=\delta _{\mathbf{k},\mathbf{k}^{\prime }}\delta _{i,j}$ or $[\hat{a}_{i,
\mathbf{k}},\hat{a}_{j,\mathbf{k}^{\prime }}^{\dagger }]=\delta _{\mathbf{k}%
, \mathbf{k}^{\prime }}\delta _{i,j}$. Here $\mathbf{k}=(2\pi /L)\mathbf{n}$ is the momentum in
wave-number units, with $\mathbf{n}=(n_{x},n_{y},n_{z})$ and $n_{i}=0,\pm
1,\pm 2,\ldots $ ($i=x,y,z$) in the 3D case, with similar relations in the
1D and 2D cases.

The Hamiltonian of the system, Eq.~(\ref{eq:Ham}), can now be written as
\begin{eqnarray}
\hat{H} &=&\sum\limits_{\mathbf{k}}\frac{\hbar ^{2}\mathbf{k}^{2}}{2m_{0}}%
\hat{b}_{\mathbf{k}}^{\dagger }\hat{b}_{\mathbf{k}}+\sum\limits_{\mathbf{k}%
,j=1,2}\left( \frac{\hbar ^{2}\mathbf{k}^{2}}{2m_{j}}+\hbar \Delta \right)
\hat{a}_{j,\mathbf{k}}^{\dagger }\hat{a}_{j,\mathbf{k}}  \notag \\
&&-i\frac{\hbar \chi _{D}}{L^{D/2}}\sum\limits_{\mathbf{q},\mathbf{k}}(\hat{b%
}_{\mathbf{q}}^{\dagger }\hat{a}_{1,\mathbf{k}}\hat{a}_{2,\mathbf{q}-\mathbf{%
\ k}}-\hat{a}_{2,\mathbf{q}-\mathbf{k}}^{\dagger }\hat{a}_{1,\mathbf{k}%
}^{\dagger }\hat{b}_{\mathbf{q}}).  \label{H-kq}
\end{eqnarray}

To proceed, we next assume that the initial condition is a large molecular
condensate in the mode $\hat{b}_{0}$ ($\langle \hat{b}_{0}^{\dagger }(0)\hat{%
b}_{0}(0)\rangle \gg 1$) and that the dynamics of dissociation retains the
molecular population in the same condensate mode $\hat{b}_{0}(t)$. This
implies that we ignore any atom-atom recombination process in which two
atoms with non-opposite momenta $\hat{a}_{1,\mathbf{k}}$ and $\hat{a}_{2,%
\mathbf{q}-\mathbf{k}}$ combine to populate a non-condensate molecular mode $%
\hat{b}_{\mathbf{q}}$ with $\mathbf{q}\neq 0$. This is a reasonable
approximation at least during the initial stages of dissociation when
recombination is negligible. Thus, the Hamiltonian (\ref{H-kq}) now reduces
to
\begin{eqnarray}
\hat{H} &=&\sum\nolimits_{\mathbf{k}}\hbar \Delta _{\mathbf{k}}\left( \hat{a}%
_{1,\mathbf{k}}^{\dagger }\hat{a}_{1,\mathbf{k}}+\hat{a}_{2,\mathbf{k}%
}^{\dagger }\hat{a}_{2,\mathbf{k}}\right)  \notag \\
&&-i\hbar \kappa \sum\nolimits_{\mathbf{k}}(\hat{b}_{0}^{\dagger }\hat{a}_{1,%
\mathbf{k}}\hat{a}_{2,-\mathbf{k}}-\hat{a}_{2,-\mathbf{k}}^{\dagger }\hat{a}%
_{1,\mathbf{k}}^{\dagger }\hat{b}_{0}),  \label{H-k}
\end{eqnarray}%
where we have introduced $\Delta _{\mathbf{k}}\equiv \Delta +\hbar \mathbf{k}%
^{2}/(2m_{1})$ and a new coupling constant%
\begin{equation}
\kappa \equiv \chi _{D}/L^{D/2}.
\end{equation}%
We note that the Hamiltonian (\ref{H-k}) conserves the total number of
atomic particles, $\hat{N}=2\hat{b}_{0}^{\dagger }\hat{b}_{0}+\sum\nolimits_{
\mathbf{k}}\left( \hat{n}_{1,\mathbf{k}}+\hat{n}_{2,\mathbf{k}}\right) $,
and the number difference, $\hat{n}_{1,\mathbf{k}}-\hat{n}_{2,\mathbf{k}}$,
where $\hat{n}_{j,\mathbf{k}}=\hat{a}_{j,\mathbf{k}}^{\dagger }\hat{a}_{j,
\mathbf{k}}$ ($j=1,2$) are the particle number operators for the atoms.

The next step in the mean-field approach is to assume that the molecular
condensate is initially in a coherent state with an amplitude $\beta _{0}$ ($%
\langle \hat{b}_{0}^{\dagger }(0)\hat{b}_{0}(0)\rangle =|\beta _{0}|^{2}\gg
1 $), and that it remains in a coherent state during the dynamical
evolution. The condensate dynamics are then treated at the level of the
mean-field amplitude, $\beta (t)=\langle \hat{b}_{0}(t)\rangle $, which
implies that we can approximate the higher-order correlation function
 $\langle \hat{b}_{0}^{\dagger }\hat{a}_{1,\mathbf{k}}\hat{
a}_{2,-\mathbf{k}}\rangle \approx \beta _{0}^{\ast }\langle \hat{a}_{1,\mathbf{k}}
\hat{a}_{2,-\mathbf{k}}\rangle $. Thus, the decorrelation assumption is
imposed at all times, which is the main limitation of the method.
This treatment will become increasingly inadequate as the
molecular population depletes and approaches zero.

Once we impose the decorrelation assumption, the dynamics of the
atomic fields can be described in terms of the normal and anomalous
populations
\begin{eqnarray}
n_{\mathbf{k}}(t) &\equiv &\langle \hat{n}_{1,\mathbf{k}}(t)\rangle =\langle
\hat{n}_{2,\mathbf{k}}(t)\rangle ,  \label{nk-def} \\
m_{\mathbf{k}}(t) &\equiv &\langle \hat{m}_{\mathbf{k}}(t)\rangle
=\langle \hat{a}_{1,\mathbf{k}}(t)\hat{a}_{2,-\mathbf{k}}(t)\rangle
, \label{mk-def}
\end{eqnarray}%
which describe the occupation numbers of the atomic modes with
momenta $\mathbf{k}$, Eq.~(\ref{nk-def}), and the correlation
between the modes with equal but opposite momenta, $\mathbf{k}$ and
$-\mathbf{k}$, in the two spin states, Eq.~(\ref{mk-def}). In this
description, higher-order correlation functions factorize according
to Wick's theorem, i.e., they can be expressed in terms of the
second-order moments $n_{\mathbf{k}}(t)$ and $m_{\mathbf{k}}(t)$.

Writing down the Heisenberg operator equations of motion following from the
Hamiltonian (\ref{H-k}) and imposing the pairing mean-field decorrelation
approximation, one can show that the equations of motion for the scaled
mean-field amplitude,%
\begin{equation}
f(\tau )\equiv \beta (t)/\beta _{0},
\end{equation}
and the normal and anomalous populations $n_{\mathbf{k}}(t)$ and $m_{\mathbf{%
k}}(t)$ form a closed set and can be written as%
\begin{eqnarray}
\frac{dn_{\mathbf{k}}(\tau )}{d\tau } &=&m_{\mathbf{k}}^{\ast }(\tau )f(\tau
)+m_{\mathbf{k}}(\tau )f^{\ast }(\tau ),  \notag \\
\frac{dm_{\mathbf{k}}(\tau )}{d\tau } &=&-2i\delta _{\mathbf{k}}m_{\mathbf{k}%
}(\tau )+f(\tau )\left[ 1\pm 2n_{\mathbf{k}}(\tau )\right] ,  \label{PMFT} \\
\frac{df(\tau )}{d\tau } &=&-\frac{1}{\beta _{0}^{2}}\sum\nolimits_{\mathbf{k%
}}m_{\mathbf{k}}(\tau ).  \notag
\end{eqnarray}%
In these equations and throughout the rest of this paper, the upper (lower)
sign refers to bosonic (fermionic) atom statistics, and we have introduced
dimensionless time $\tau =t/t_{0}$ and dimensionless effective detuning
\begin{eqnarray}
\delta _{\mathbf{k}}=\Delta _{\mathbf{k }}t_{0}=[\hbar \mathbf{k}%
^{2}/(2m_{1})+\Delta ]t_{0}=q^{2}+\delta,
\end{eqnarray}
where $t_{0}=1/(\kappa \beta _{0})$ is the time scale (where we have assumed
that $\beta _{0}$ is real without loss of generality), $q=|\mathbf{q}|$ is
the absolute value of the dimensionless momentum $\mathbf{q}=\mathbf{k}d_{0}$%
, $d_{0}=\sqrt{\hbar t_{0}/(2m_{1})}$ is the length scale, and $\delta
=\Delta t_{0}$ is the dimensionless bare detuning. The quantity $f(\tau )$
is the fractional molecular amplitude so that $|f(\tau )|^{2}=N_{0}(\tau
)/N_{0}(0) $ corresponds to the fraction of molecules relative to their
initial total number. Equations (\ref{PMFT}) conserve the
quantity $|m_{\mathbf{k}}(t)|^{2}-n_{\mathbf{k}}(t)[1\pm n_{\mathbf{k}}(t)]$
as its time derivative is zero, and so the pair and anomalous densities are directly
related according to
\begin{eqnarray}
|m_{\mathbf{k}}(t)|^{2}=n_{\mathbf{k}}(t)[1\pm n_{\mathbf{k}}(t)].
\label{eq:nm}
\end{eqnarray}
This will be useful in the calculation of correlation functions in
Sec.~\ref{aa_corr}.

Equations~(\ref{PMFT}) are equivalent to those solved in Ref.~\cite{Jack-Pu}
for describing dissociation dynamics in 3D using spin-1/2 Pauli matrices
\cite{Comment1}. Here, we extend this treatment to 1D and 2D systems. In
addition, we analyze the second-order correlation functions for the atoms
and give detailed quantitative assessment of the approximations involved in
the pairing mean-field method compared to the exact first-principles
treatment using the positive $P$-representation for bosons \cite{Savage}.

\subsubsection{Results for total atom numbers}

\label{sect:total-N}

\begin{figure}[tbp]
\includegraphics[height=4.4cm]{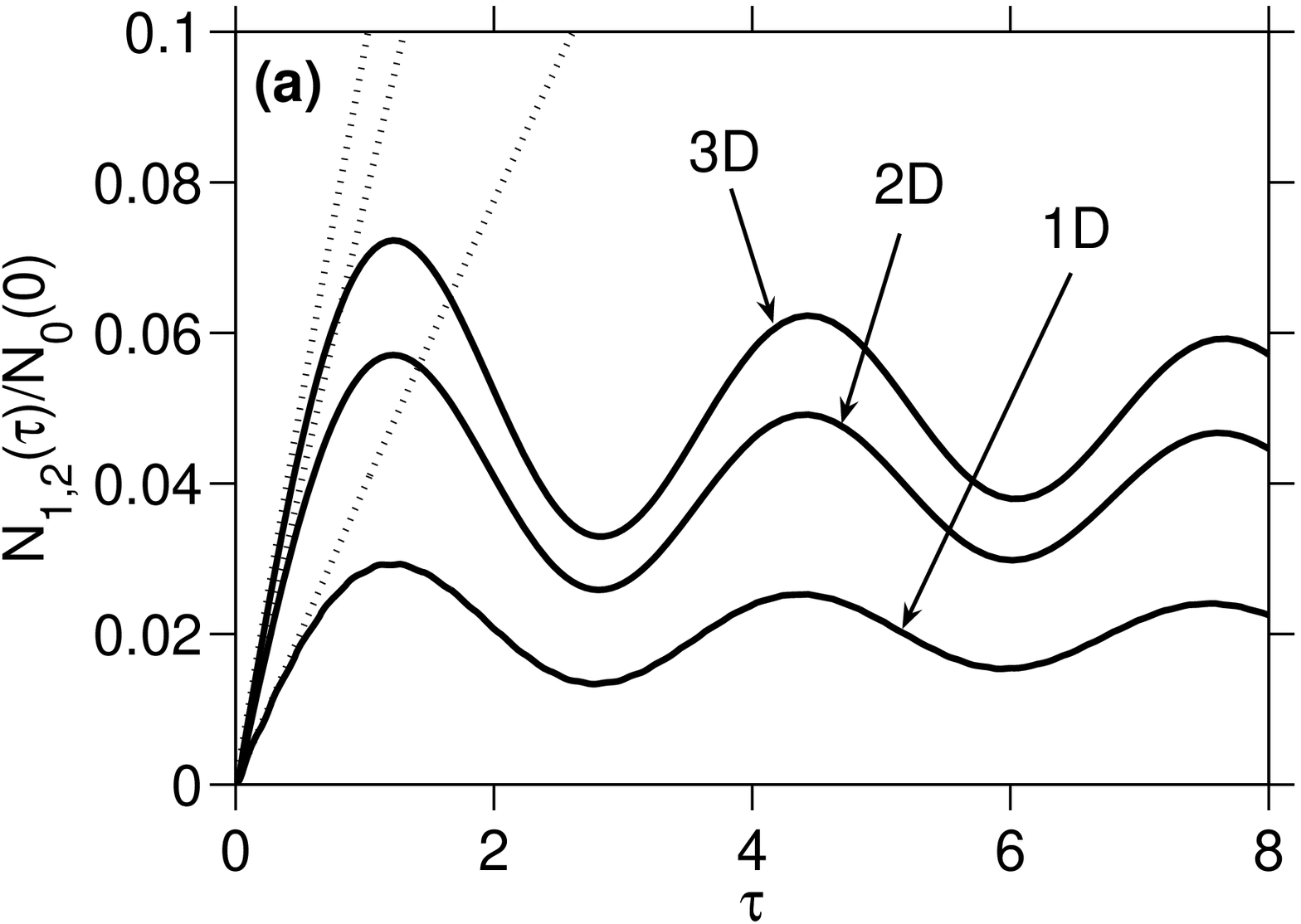}
\includegraphics[height=4.4cm]{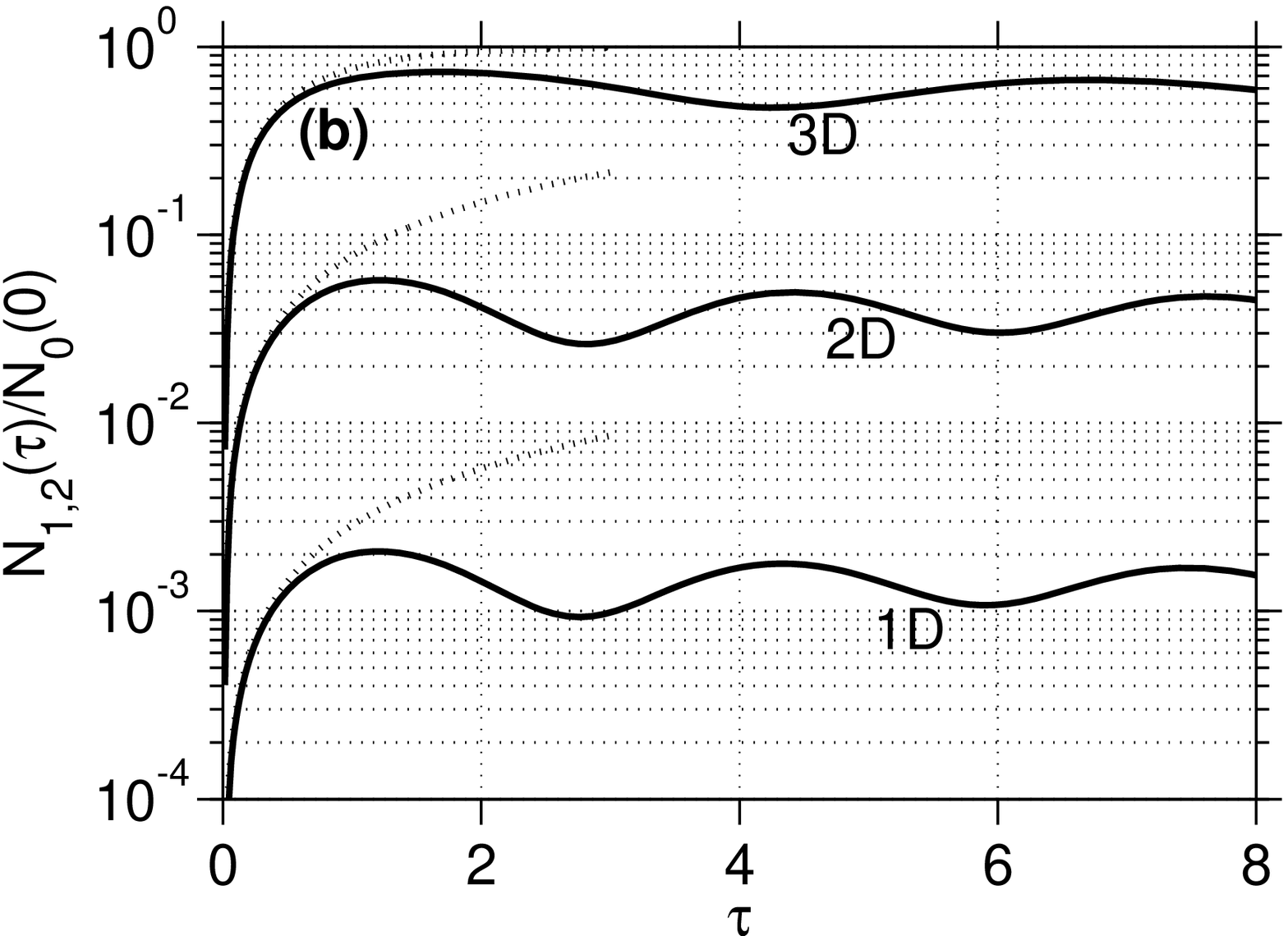}
\caption{Dissociation of a molecular condensate into fermionic atoms. We
plot the total number of atoms in each spin component $N_{1,2}(\protect\tau %
) $ [$N_{1}(\protect\tau )=N_{2}(\protect\tau )$] relative to the total
initial number of molecules $N_{0}(0)$ as a function of the dimensionless
time $\protect\tau =t/t_{0}$, where $t_{0}=1/\protect\kappa \protect\sqrt{%
N_{0}(0)}$ is the time scale. The graphs (a) and (b) are for $|\protect%
\delta |=16$ and $|\protect\delta |=3174$, respectively. The different
curves are for 1D, 2D and 3D cases as shown. In all cases, the initial
dimensionless molecular density is $N_{0}(0)/l^{D}=3.1$ ($D=1,2,3$), where $%
l=L/d_{0}$ is the dimensionless length. The dotted lines next to each solid
line are the results from Fermi's golden rule, Eqs.~(\protect\ref{Gamma})
and (\protect\ref{Atom-number}). See text for further details. }
\label{fig1}
\end{figure}

\begin{figure}[tbp]
\includegraphics[height=4.4cm]{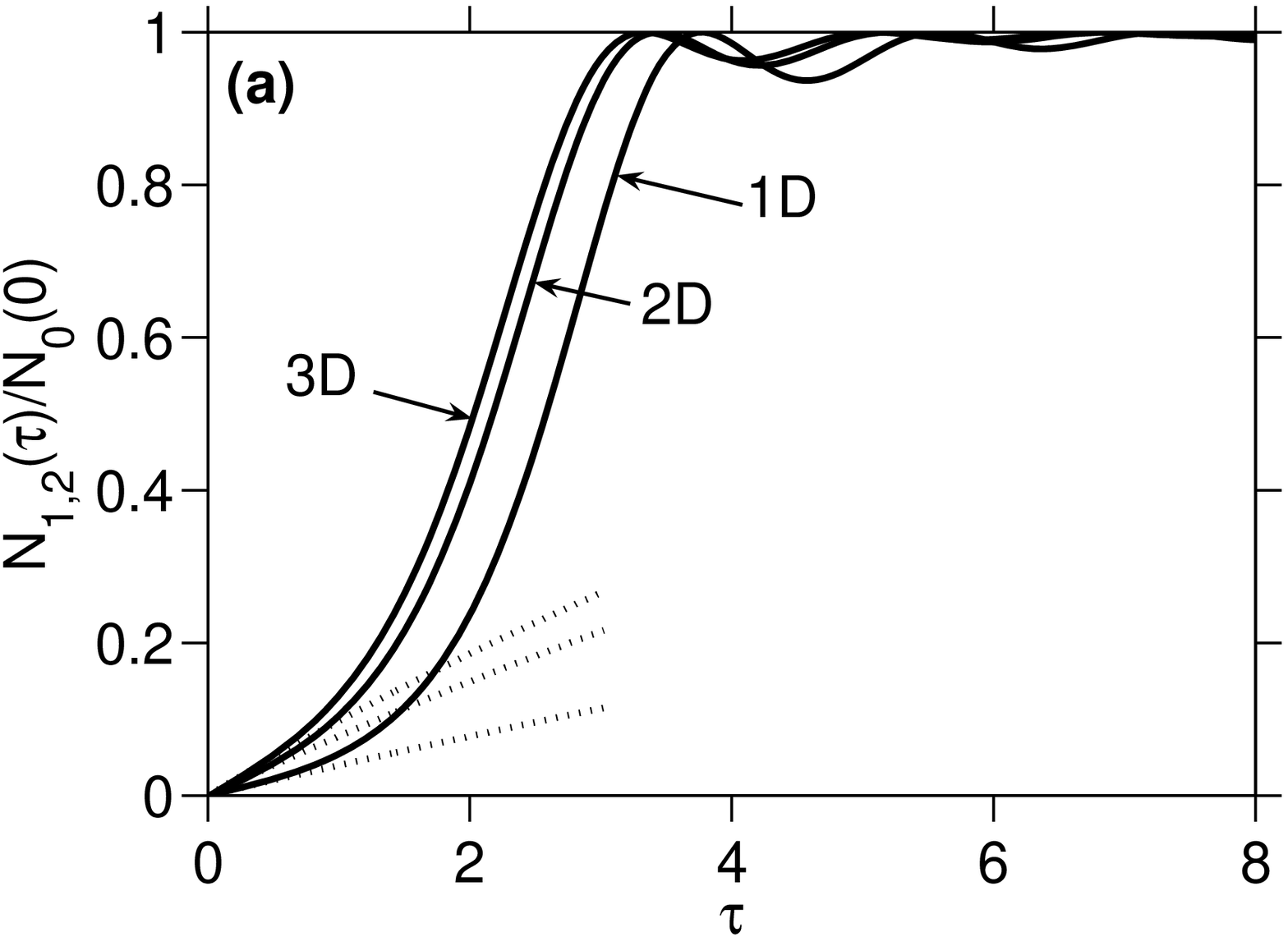}
\includegraphics[height=4.4cm]{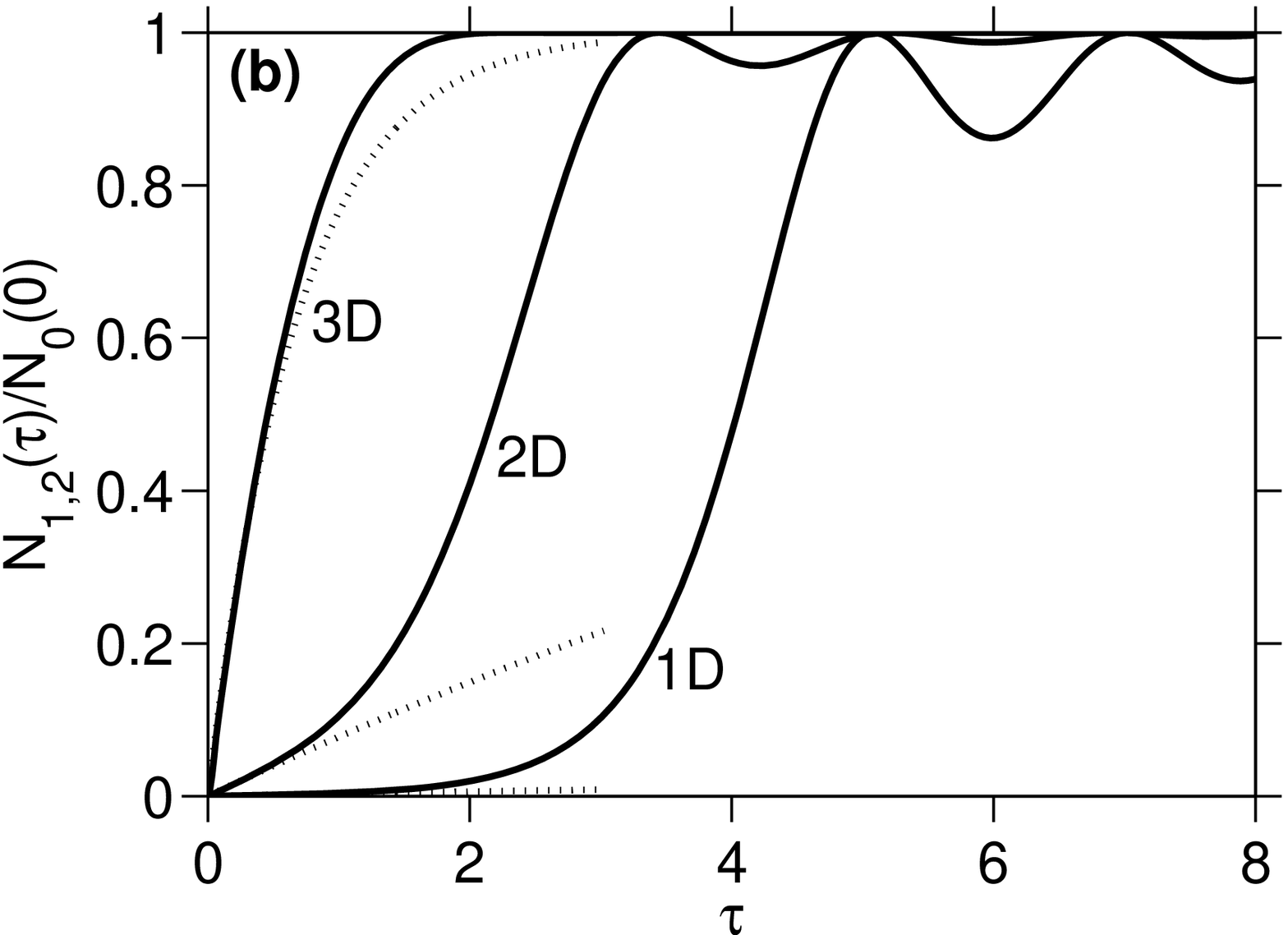}
\caption{Dissociation of a molecular condensate into distinguishable bosons.
Apart from bosonic statistics, all parameters are the same as for
Fig.~\ref{fig1}.}
\label{fig2}
\end{figure}

In Figs.~\ref{fig1} and \ref{fig2} we show typical results from solving
Eqs.~(\ref{PMFT}) for fermionic and
bosonic atom statistics respectively. We plot the fraction of the total
number of atoms in each spin component relative to the total initial number
of molecules, $N_{1,2}(\tau )/N_{0}(0)$, as a function of time. The two
graphs, (a) and (b), are for two different values of the dimensionless
detuning $\delta =\Delta t_{0}$ ($\delta <0$) and the same initial
dimensionless molecular density $N_{0}(0)/l^{D}=3.1$ ($D=1,2,3$), where $%
l=L/d_{0}$ is the dimensionless length corresponding to the quantization
length $L$, with $d_{0}=\sqrt{\hbar t_{0}/(2m_{1})}$ being the length scale.

These parameters can be size-matched with a realistic nonuniform system
corresponding, for example, to a BEC of molecular dimers made of fermionic $%
^{40}$K atoms as in Ref.~\cite{Greiner}. Our empirically derived
prescription for the size-matching \cite{Savage} is as follows.
Assuming for simplicity an isotropic harmonically trapped molecular
condensate with a Thomas-Fermi parabolic density profile, we match
the parameters of the present uniform treatment to have the same
peak density $\rho _{0}$ as the trapped condensate and choose the
length $L$ of the uniform box to result in the same initial total
number of molecules as in the trapped condensate $N_{0}(0)=(8\pi
/15)R_{TF}^{3}\rho _{0}=L^{3}\rho _{0}$, where $R_{TF}$ is the
Thomas-Fermi radius. Thus, the length $L$ is size-matched to
$L^{3}=(8\pi /15)R_{TF}^{3}$ \cite{Fermidiss,Savage}. Taking the
peak density $\rho _{0}=7.6\times 10^{18} $ m$^{-3}$, the trap
frequency $\omega /2\pi =52$ Hz, and molecule-molecule scattering
length $a_{mm}\simeq 1.47$ $\mu $m in the
strongly interacting regime near a magnetic Feshbach resonance, we obtain $%
R_{TF}=\sqrt{8\pi \hbar ^{2}a_{mm}\rho _{0}/m_{0}^{2}\omega ^{2}}\simeq 33.1$
$\mu $m, $L\simeq 39.3$ $\mu $m, and a total number of molecules $%
N_{0}(0)\simeq 3\times 10^{5}$. Assuming a 3D atom-molecule
coupling $\chi _{3D}=4.87\times 10^{-7}$ m$^{3/2}$s$^{-1}$, we obtain a time
scale of $t_{0}=1/[\kappa \sqrt{N_{0}(0)}]\simeq 0.918$ ms, where $\kappa
=\chi _{3D}/L^{3/2}\simeq 1.98$ s$^{-1}$ and giving a length
scale of $d_{0}\simeq 0.854$ $\mu $m, $l=L/d_{0}\simeq 46$, and hence $%
N_{0}(0)/l^{3}\simeq 3.1$.

Most of the numerical examples given here are for two values of the
dimensionless detuning, $|\delta |=16$ and $|\delta |=3174$, where
$\delta =\Delta t_{0}$. Using the above timescale of $t_{0}\simeq
0.918$ ms, these detunings correspond to $|\Delta |/2\pi \simeq
3.67$ kHz and $|\Delta |/2\pi \simeq 0.55$\ MHz. The second case is
chosen to coincide with the average rf detuning of $\nu _{rf}=1.1$
MHz used in the experiment of Ref.~\cite{Greiner}, where we note
that $\nu_{rf}$ corresponds to the absolute detuning of $|\Delta
|/\pi $ in our notation. The duration of dissociation of $0.33$ ms
used in Ref.~\cite{Greiner} corresponds to dimensionless time of
$\tau \simeq 0.36$ at which
 our 3D results give $N_{1}(t)/N_{0}(0)\simeq 0.39$ ($39\%$
conversion), in good agreement with $\sim 43\%$ conversion
obtained in Ref.~\cite{Greiner}. We also note that the dimensionless
time $\tau _{\max }=l/(2\sqrt{|\delta |})$ corresponding to Eq.
(\ref{box-time}) is equal to $\tau _{\max }\simeq 0.41$ in this
example, which is larger than the dissociation duration of $\tau
\simeq 0.36$ as required for the applicability of the uniform
treatment.

More detailed quantitative comparison with the results of Ref.
\cite{Greiner}, especially those for the momentum distribution and
atom-atom correlations, is complicated by the fact that the detuning
$\nu _{rf}$ was swept over $600$ kHz during the experiment in order
to spread the detected signal over many camera pixels. We do not
model the rf sweep in the present work. Additional discrepancies are
expected to arise due to the fact that the trap geometry in the
experiment was not spherically symmetric as assumed here.

The results of Figs.~\ref{fig1} and \ref{fig2} are easily understood using
the dependence of the density of atomic states on energy, Eq.~(\ref%
{densityofstates}). By energy and momentum conservation, the total
dissociation energy $2\hbar |\Delta |$ is converted into the kinetic
energy, $2\hbar |\Delta |\rightarrow \hbar ^{2}k_{\uparrow
}^{2}/(2m_{1})+\hbar ^{2}k_{\downarrow }^{2}/(2m_{1})$, of atom
pairs with opposite spins and momenta. The atomic momenta are
distributed in a narrow interval around $k_{0}=|\mathbf{k}_{\uparrow
(\downarrow )}|= \sqrt{2m_{1}|\Delta |/\hbar }$. In 3D, the resonant
momenta form a spherical shell with a radius $k_{0}$; in 2D they are
distributed within a circular shell at the same radius, while in 1D
the resonant momenta are around $\pm k_{0}$. Since the density of
atomic states at the dissociation energy is the largest in 3D, the
number of dissociated atoms is larger in 3D than in 2D and 1D.

Next, we address the question of how the number of dissociated atoms
changes with the dissociation energy $2\hbar|\Delta |$. Since the
density of states increases with energy in 3D, a larger absolute
detuning results in a larger number of atoms produced [compare the
3D curves in Figs. \ref{fig1}(a) and (b); also in Figs.
\ref{fig2}(a) and (b)]. In 1D the situation is reversed; the density
of states decreases with energy and hence the number of atoms at a
given time is smaller at larger absolute detuning $|\Delta |$. In
2D, the density of states is independent of the energy and we see no
variation of the total
atom number with the detuning. Indeed, the 2D curves in Figs.~\ref{fig1} and %
\ref{fig2} are indistinguishable within numerical accuracy, even though the
equations were solved with different detunings.

As expected, the dependencies on the absolute detuning in 1D, 2D and
3D found numerically are in agreement with the explicit analytic
results for the atom production rate found from Fermi's golden rule,
Eq. (\ref{Atom-number}). The atom numbers from the simple rate
equation (\ref{Atom-number}) are shown by dotted lines in Figs.
\ref{fig1} and \ref{fig2}, where they appear as the tangents to the
respective numerical results at $\tau \rightarrow 0$.

Comparing the results of the pairing mean-field theory with those based on
Fermi's golden rule calculations of the molecular decay rate, we see that in
the initial spontaneous stage of dissociation the bosonic and fermionic
results are very similar. Past the spontaneous regime, the dynamics of
dissociation becomes affected by either Pauli blocking or bosonic
stimulation, depending on the statistics of the dissociated atoms.
Accordingly, the fermionic results for the total atom number saturate faster
and lie below the curve corresponding to the simple rate equation, while the
bosonic results exhibit exponential growth due to Bose enhancement and
remain above the rate equation curve.

\begin{figure}[tbp]
\includegraphics[height=4.4cm]{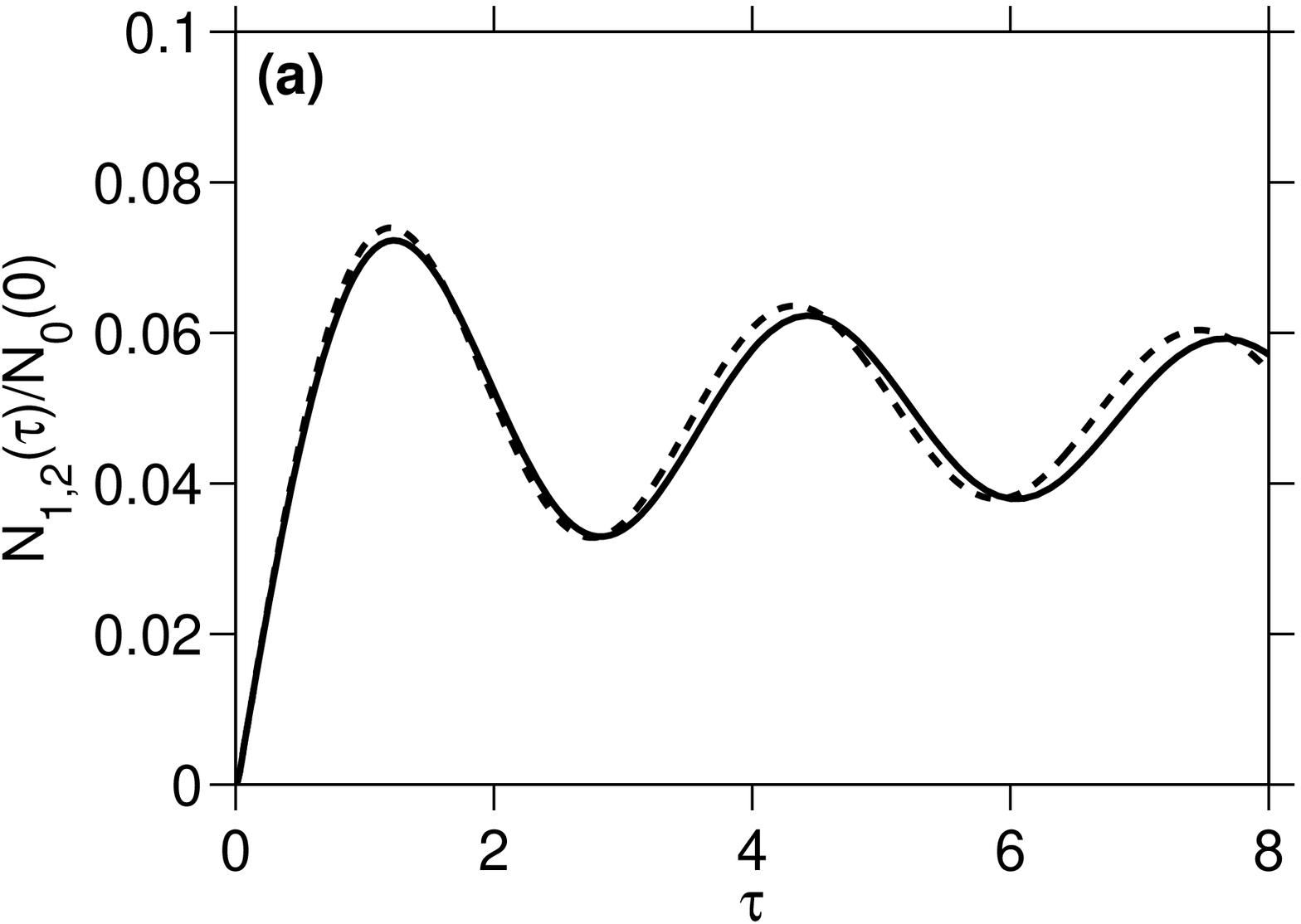}
\includegraphics[height=4.4cm]{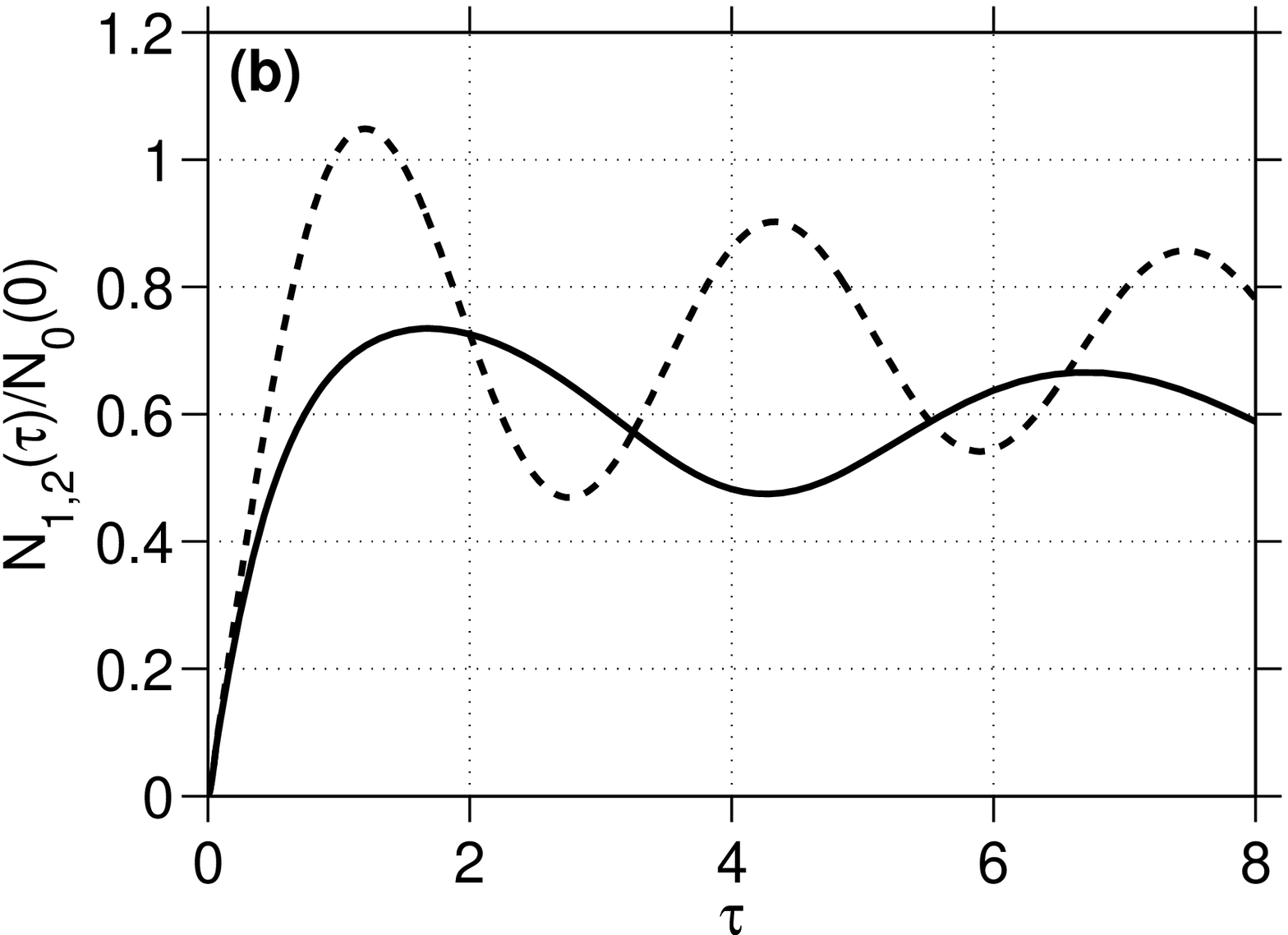}
\caption{Same as in Fig. \protect\ref{fig1}(a) and (b) in the 3D case,
except that the results of the pairing mean-field theory (solid lines) are
compared with those obtained within the undepleted molecular field
approximation (dashed line). }
\label{fig3}
\end{figure}

In the bosonic case, the ultimate limit to the growth in atom
number is set by the total (finite) initial number of molecules as
the condensate can be entirely depleted. For fermions, however, the
Pauli exclusion principle -- depending on dimensionality of space
and absolute detuning -- can take effect before there is any
significant depletion of the molecular condensate. In this sense the
molecular depletion is a more important factor in the bosonic case
than in the fermionic case. This situation applies to
all curves in Fig.~\ref{fig1}(a) and to 1D and 2D results in Fig.~\ref{fig1}%
(b). In all these fermionic cases the conversion efficiency is less
than $8\%$ during the entire simulation time and the molecular
depletion is negligible. Accordingly, the dynamics of dissociation
follows closely the predictions obtained within the undepleted
molecular field approximation of Ref.~\cite{Fermidiss}. This is
shown in Fig.~\ref{fig3}(a), where we present the comparison in the
3D case, where the discrepancy is the largest. Owing to the
different dependencies of the density of states on the absolute
detuning in different dimensions, the undepleted molecular field
approximation works better at small absolute detunings in 3D and at
large detunings in 1D. For the same reason, the molecular depletion
generally is less important in 1D than in 3D.

If, on the other hand, the number of available atomic states in the
spherical shell around $k_{0}$ [see Eq.~(\ref{width-stim}) for the
width of the shell] is comparable to or larger than the total
initial number of molecules, then the dynamics of dissociation is
dominated by the molecular field depletion. In this case, the number
of molecules can decrease substantially before the population of
individual atomic modes experiences any Pauli blocking. This is a
typical situation in 3D at very large absolute detunings $|\delta |$
when there is a large number of states available for occupancy. Even
if the average occupation of each of these states is smaller than
one, the total average number of atoms can be quite large and
constitute a large fraction of the initial number of molecules.
Accordingly, the predictions of the undepleted molecular field
approximation remain valid only for very short time as seen in
Fig.~\ref{fig3}(b). In the longer time limit, the undepleted
molecular field approximation leads to an unphysical result that the
fractional atomic population becomes larger than one, as seen in
Fig.~\ref{fig3}(b).
 This situation is similar to the well-known behavior in the bosonic
case (see Fig.~\ref{fig5} below).


\subsubsection{Parametrization at large dissociation energy}

Once the equations of motion are written in  dimensionless form it is clear
the system can be described via just two parameters --- the
dimensionless detuning $\delta $ and the initial number of molecules $%
N_{0}=\beta _{0}^{2}$.  The quantization volume drops out of the final results if we are interested in
fractional populations or normalized densities rather than their absolute
values.

For large dissociation energy (large absolute detuning, $|\delta
|\gg 1$), the parametrization of the system can be further
simplified and reduced to just one parameter which is a function of
$N_{0}$ and $|\delta |$ \cite{Jack-Pu}. Here, we show this for the
case of distinguishable atoms and note that the same
arguments apply to the indistinguishable case. As shown in Appendix \ref%
{sect:appendix-C}, the parameter in question originates from the approximate
form of Eqs. (\ref{PMFT}) in the continuous limit, which gives the following
equation for the fractional amplitude%
\begin{equation}
\frac{df(\tau )}{d\tau }\simeq -\sqrt{\frac{2}{\Upsilon }}\int_{-\infty
}^{\infty }d\delta _{k}\,m(\delta _{k},\tau ).
\end{equation}%
Here, the dimensionless parameter $\Upsilon $ is defined via
\begin{equation}
\Upsilon =\frac{2N_{0}}{\hbar ^{2}\kappa ^{2}[D(\hbar |\Delta |)]^{2}},
\end{equation}%
and coincides with the parameter $\Gamma $ introduced in Ref. \cite{Jack-Pu}
(where we note that $N=2N_{0}$); this is the only parameter that
characterizes the system at large detunings. Using Eq. (\ref{densityofstates}%
), $\Upsilon $ can be written down explicitly for 1D, 2D, and 3D as follows:
\begin{equation}
\Upsilon =\left\{
\begin{array}{ll}
8\pi ^{2}|\delta |N_{0}^{2}/l^{2},&\mathrm{(1D)}, \\
\\
32\pi ^{2}N_{0}^{2}/l^{4},&\mathrm{(2D)}, \\
\\
32\pi ^{4}N_{0}^{2}/(l^{6}|\delta |),&\mathrm{(3D)}.%
\end{array}%
\right.  \label{Gamma-Jack-Pu}
\end{equation}

This approximation breaks down for small absolute detunings where
the dissociation predominantly populates a range of atomic momenta
close to zero and the density of states here can no longer be
approximated as flat. As a result the system is parameterized by two
variables $\Upsilon$ and $|\delta |$, or equivalently in terms of
the original pair $N_{0}$ and $|\delta |$, without the need to
introduce $\Upsilon $.
\begin{figure}[tbp]
\includegraphics[height=4.4cm]{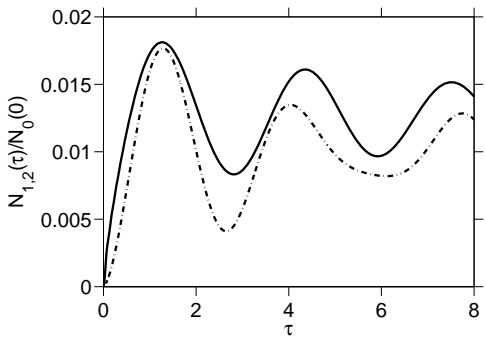}
\caption{Comparison of the results obtained without (solid line) and with
(dash-dotted line) the flat density-of-states approximation in 3D. In both
cases we set $|\protect\delta |=1$ and $N_{0}/l^{3}=3.1$, corresponding to $%
\Upsilon =3\times 10^{4}$. For the dash-dotted line, only the value of $%
\Upsilon $ matters, and the same curve\ can be obtained with different
individual values of $|\protect\delta |$ and $N_{0}/l^{3}$ as long as they
result in the same $\Upsilon $, Eq.~(\protect\ref{Gamma-Jack-Pu}).}
\label{fig4}
\end{figure}

In Fig.~\ref{fig4} we show the comparison between the results obtained with
and without the use of the flat density of states approximation. In this
example, the absolute dimensionless detuning is $|\delta |=1$ and we
see significant discrepancy between the two curves. The discrepancy
increases further for smaller detunings, while it becomes negligible for
detunings $|\delta |\gtrsim 16$.


\subsection{Indistinguishable atoms}

\label{exponential}

In the case of dissociation into indistinguishable atoms our
treatment corresponds to the Hamiltonian
\begin{eqnarray}
\hat{H} &=&\sum\nolimits_{\mathbf{k}}\hbar \Delta _{\mathbf{k}}\hat{a}_{1,%
\mathbf{k}}^{\dagger }\hat{a}_{1,\mathbf{k}}  \notag \\
&&-i\frac{\hbar \kappa }{2}\sum\nolimits_{\mathbf{k}}(\hat{b}_{0}^{\dagger }%
\hat{a}_{1,\mathbf{k}}\hat{a}_{1,-\mathbf{k}}-\hat{a}_{1,-\mathbf{k}%
}^{\dagger }\hat{a}_{1,\mathbf{k}}^{\dagger }\hat{b}_{0}),
\end{eqnarray}%
where\ again $\kappa \equiv \chi _{D}/L^{D/2}$ and $\chi _{D}$ ($D=1,2,3$)
are given by Eqs. (\ref{chi3D-homonuclear}), (\ref{chi-1D}), and (\ref%
{chi-2D})

The analysis of this system within the PMFT is essentially the same
as in the distinguishable case, Eq. (\ref{PMFT}). The only
difference is in the equation for $f(\tau )$, which now reads as
\begin{equation}
\frac{df(\tau )}{d\tau }=-\frac{1}{2\beta _{0}^{2}}\sum\nolimits_{\mathbf{k}%
}m_{\mathbf{k}}(\tau ),  \label{PMFT-samespin}
\end{equation}%
while the normal and anomalous populations are%
\begin{eqnarray}
n_{\mathbf{k}}(t) &\equiv &\langle \hat{n}_{1,\mathbf{k}}(t)\rangle =\langle
\hat{a}_{1,\mathbf{k}}^{\dagger }(t)\hat{a}_{1,\mathbf{k}}(t)\rangle , \\
m_{\mathbf{k}}(t) &\equiv &\langle \hat{m}_{\mathbf{k}}(t)\rangle =\langle
\hat{a}_{1,\mathbf{k}}(t)\hat{a}_{1,-\mathbf{k}}(t)\rangle .
\end{eqnarray}

\begin{figure}[tbp]
\includegraphics[height=4.4cm]{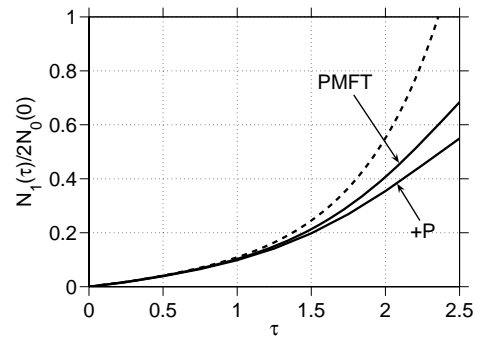}
\caption{Fractional atomic population $N_{1}(\protect\tau )/2N_{0}(0)$ in 3D
obtained using the present pairing mean-field theory (PMFT) and
first-principles (exact) simulations of the same system using the positive-$%
P$ representation method ($+P$) \protect\cite{Savage}. The dashed
line corresponds to the undepleted molecular field approximation.
The parameter values are: $|\protect\delta |=4$ and
$N_{0}(0)/l^{3}=0.988$. } \label{fig5}
\end{figure}

In Fig. \ref{fig5} we show a comparison between the present pairing
mean-field theory and first-principles (exact) simulations of the same
system using the positive-$P$ representation method \cite{Savage}. As we
see, the pairing mean-field theory compares well with the exact results for
dissociation durations corresponding to less than $50\%$ conversion at which
stage the discrepancy reaches $\sim 20\%$. This is much better than the
results obtained using the undepleted molecular field approximation \cite%
{Savage}, shown by the dashed line.

We note, however, that the comparison between the PMFT results and
those of the positive-$P$ method are not entirely equivalent as the
positive-$P$ method takes into account the possibility of a
dynamical population of the initially unoccupied molecular modes
with non-zero momenta. As has been shown in Ref.~\cite{Savage}, this
process becomes a sizeable effect as  time
progresses and occurs due to ``rogue'' association in which a small
fraction of newly formed atoms convert back into molecules with
nonzero momenta. The present version of the pairing mean-field
scheme only treats the molecular condensate mode $\hat{b}_{0}$ [see
Eq.~(\ref{H-k})] without allowing for population of the
non-condensate modes $\hat{b}_{ \mathbf{k}}$ ($\mathbf{k}\neq 0$)
present in the original Hamiltonian~(\ref{H-kq}). The performance of
the PMFT can be improved  (at the cost of increased computational
complexity) by incorporating the mean-field dynamics of all
non-condensate modes $\langle \hat{b}_{\mathbf{k}}(t)\rangle
\rightarrow \beta _{\mathbf{k}}(t)$, irrespective of their initial
population. In this manner one can extend the
present treatment from uniform to physically more realistic
nonuniform systems corresponding to spatially inhomogeneous trapped
condensates. This will be considered in future work.

\section{Momentum distribution and atom-atom correlations}

\subsection{Atomic momentum distribution}

\begin{figure}[tbp]
\hspace{0.8cm}\includegraphics[height=3.95cm]{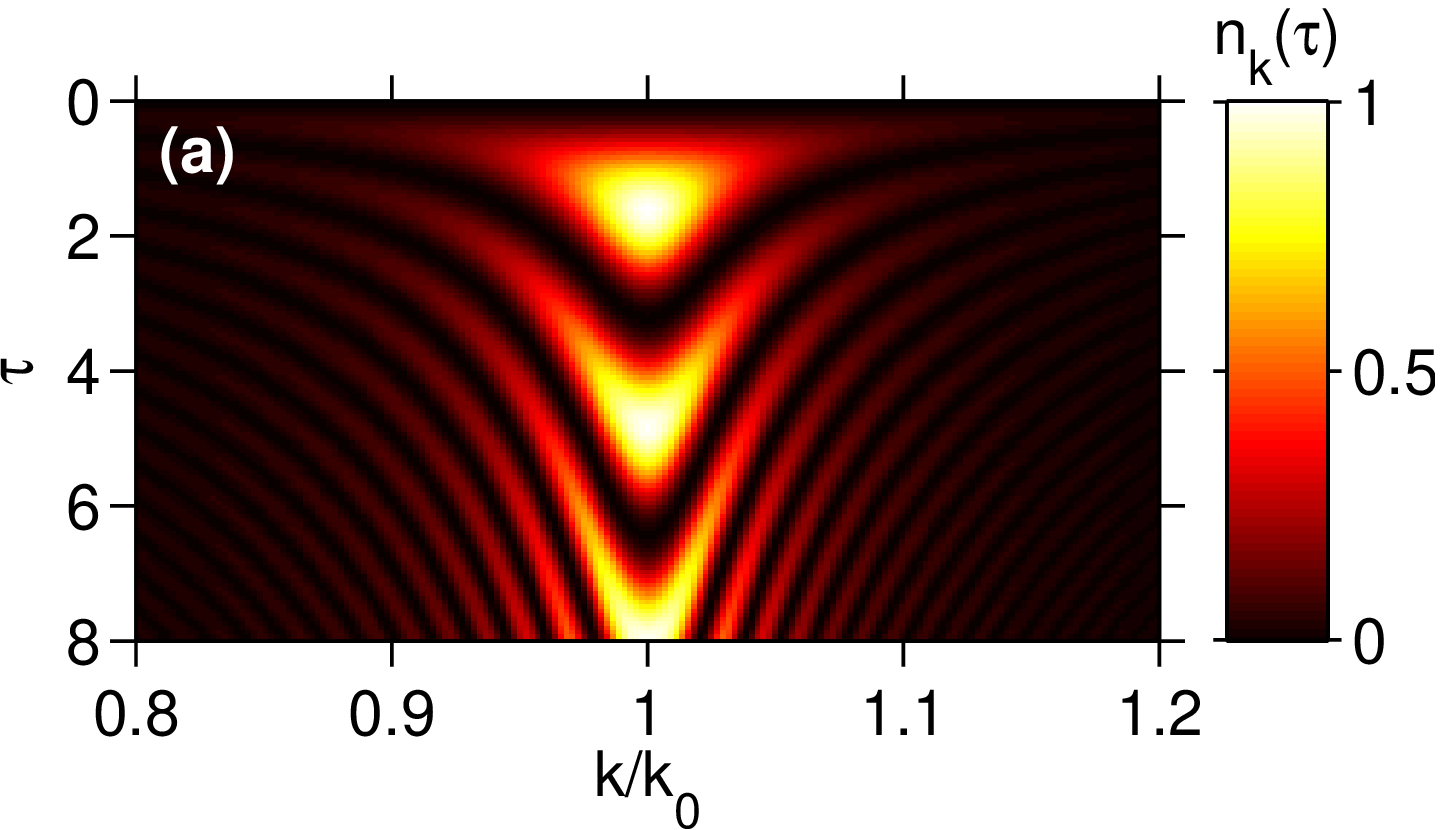}
\includegraphics[height=3.55cm]{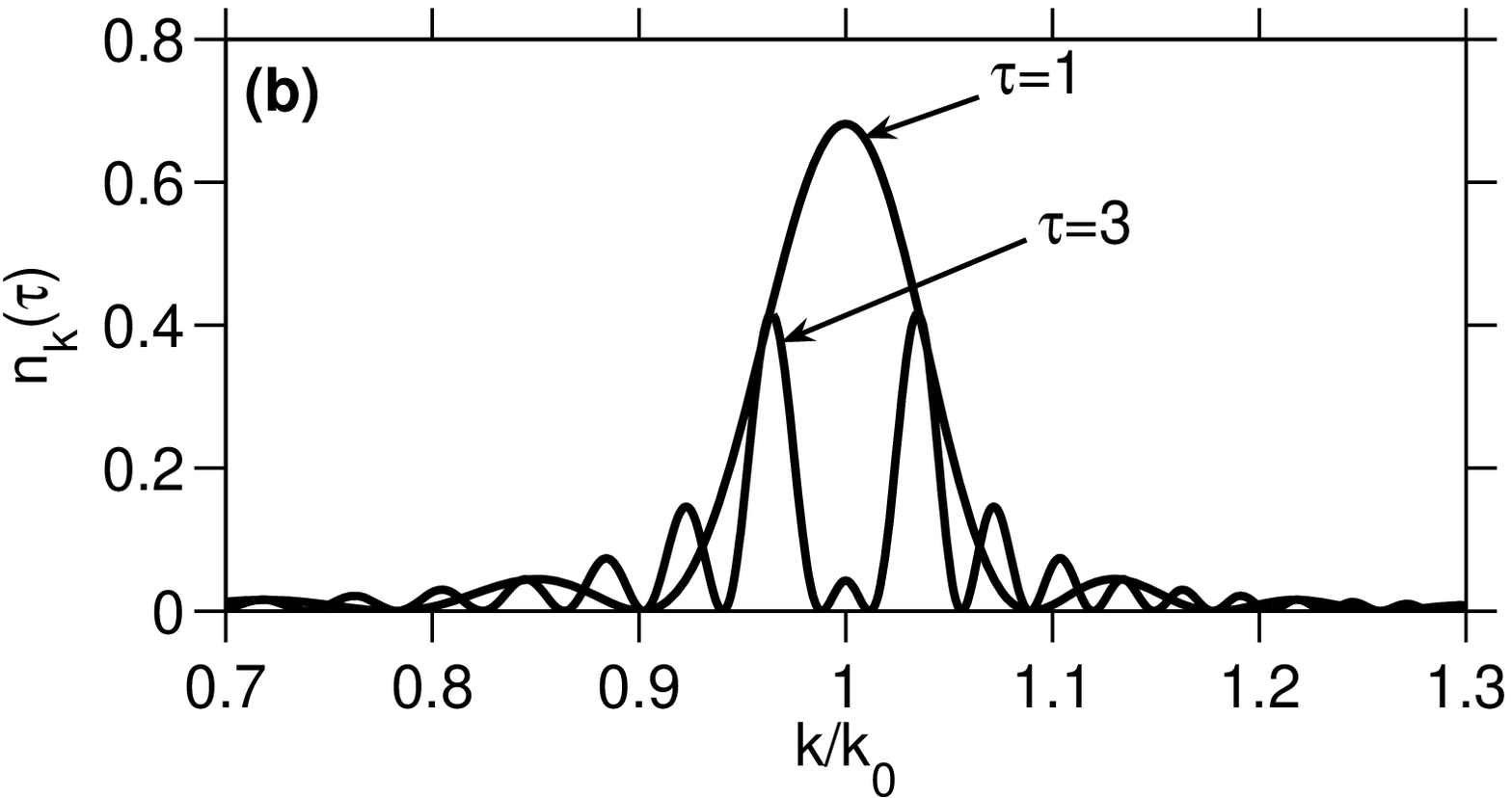}
\includegraphics[height=3.4cm]{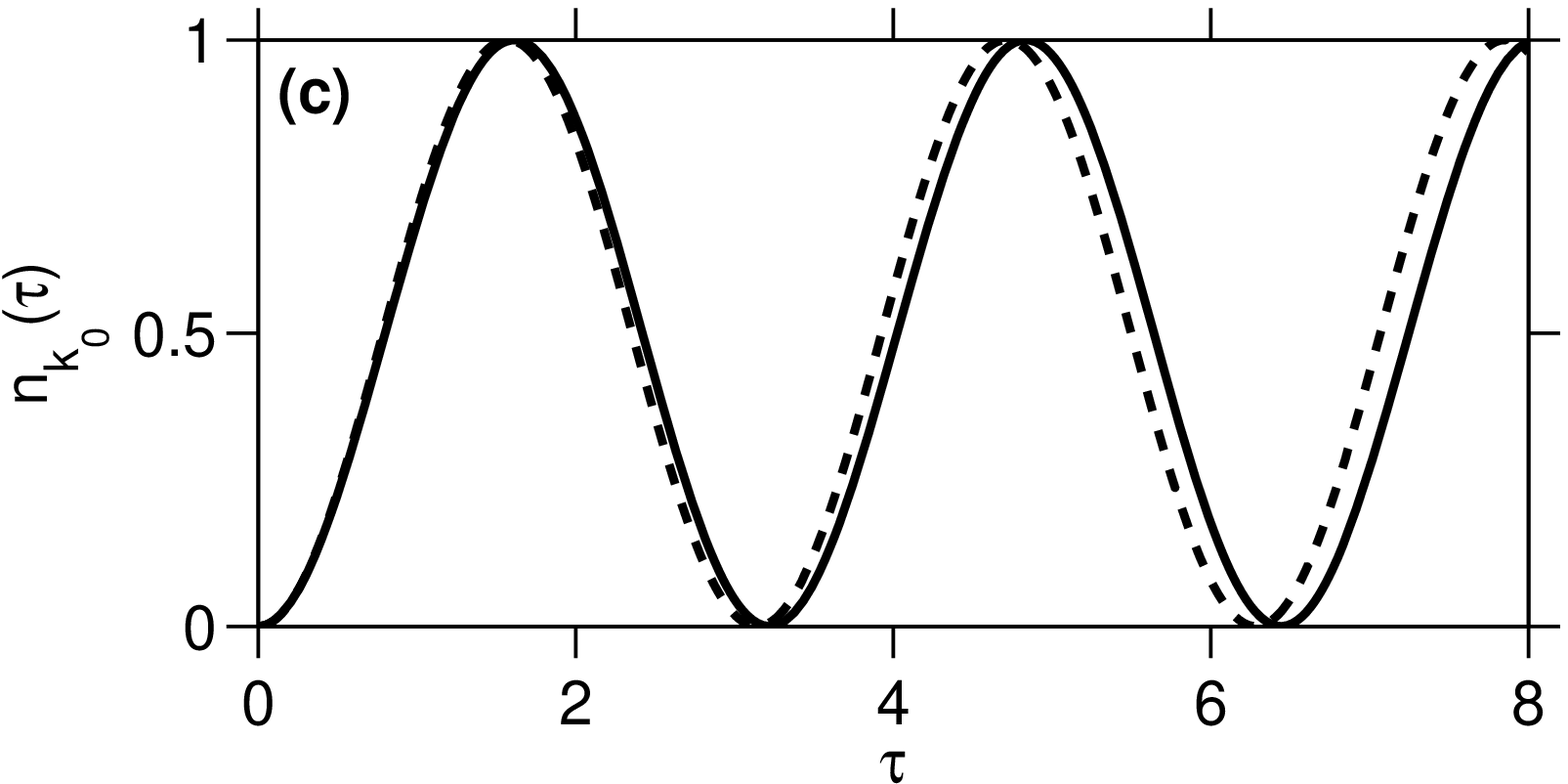}
\caption{Atomic momentum distribution in each spin state for dissociation
into fermionic atoms for a dimensionless absolute detuning $|\protect\delta %
|=|\Delta |t_{0}=16$ and $N_{0}(0)/l^{D}=3.1$. Only the range of momenta
whose population is nonvanishing is shown. (a) $n_{\mathbf{k}}(\protect\tau)$
as a function of scaled time $\protect\tau =t/t_{0}$ and scaled absolute
momentum $k/k_{0}$ in 3D. (b) Time slices of the momentum distribution at $%
\protect\tau =1$ and $\protect\tau =3$. (c) Temporal population of the
resonant momentum mode with $k=k_{0}$. The dashed line represents the
respective analytic solution given by $\sin ^{2}(\protect\tau )$ in the
undepleted molecular field approximation \protect\cite{Fermidiss}.}
\label{fig6}
\end{figure}

\begin{figure}[tbp]
\hspace{0.8cm}\includegraphics[height=3.95cm]{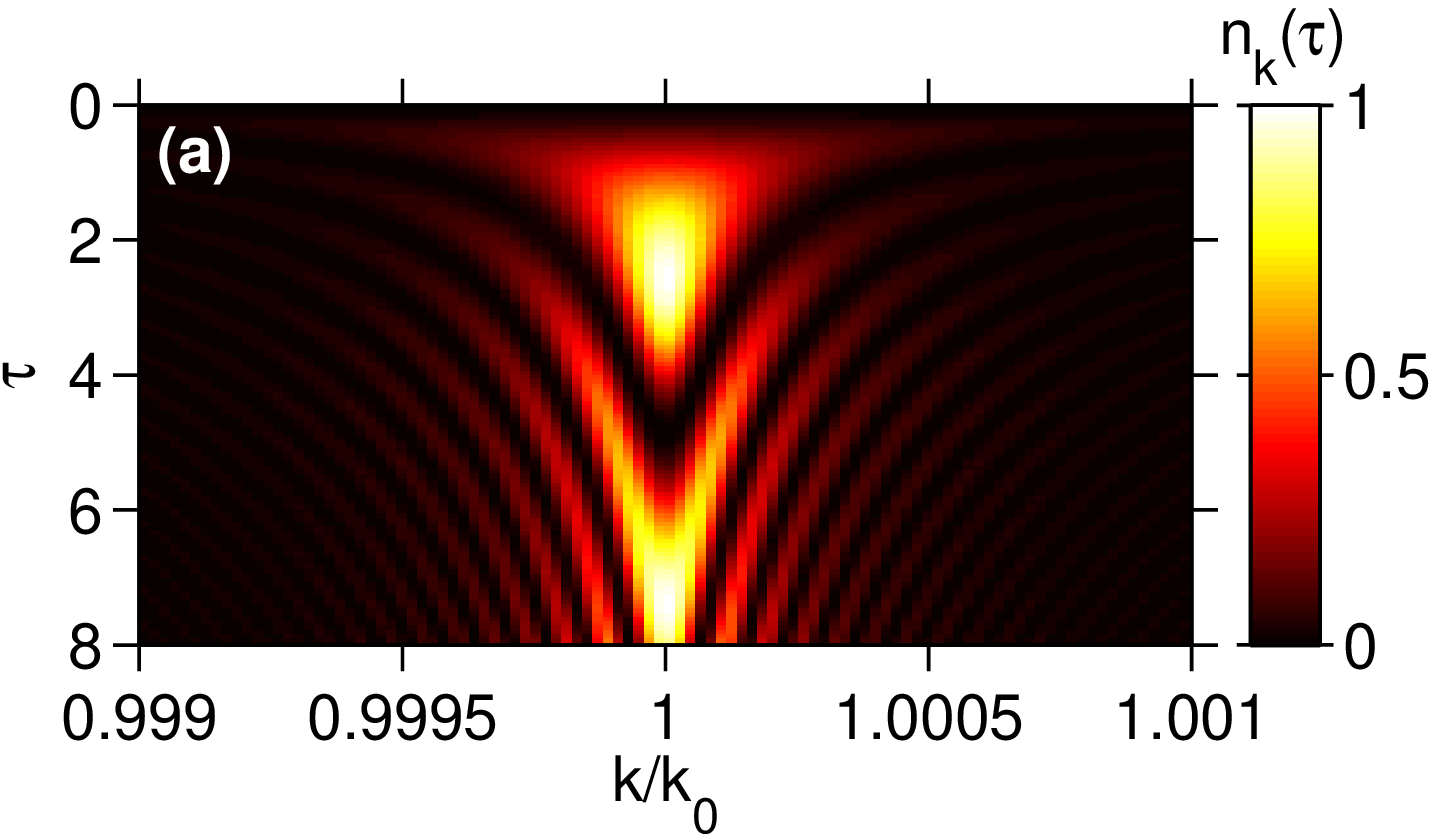}
\includegraphics[height=3.55cm]{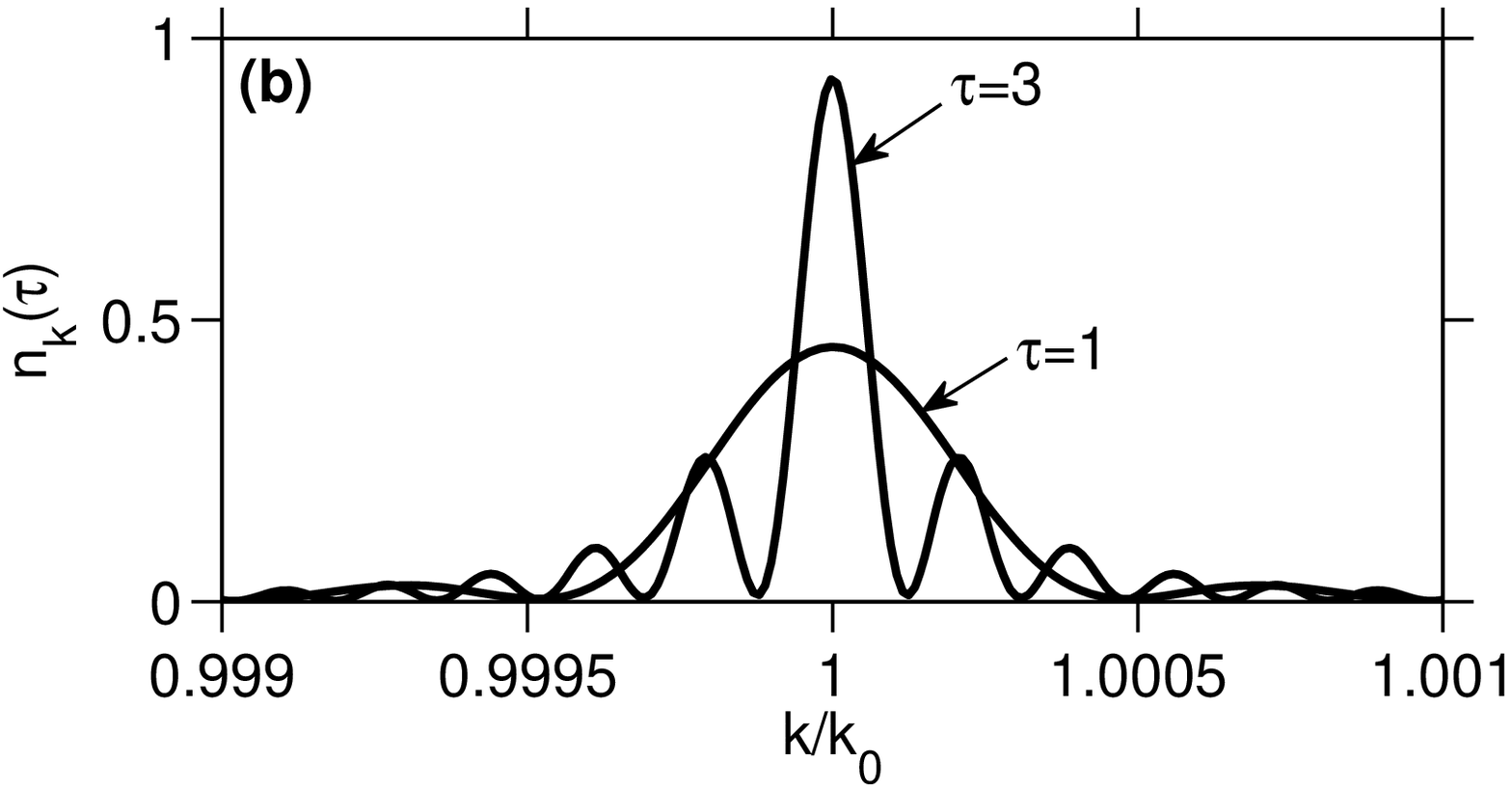}
\includegraphics[height=3.4cm]{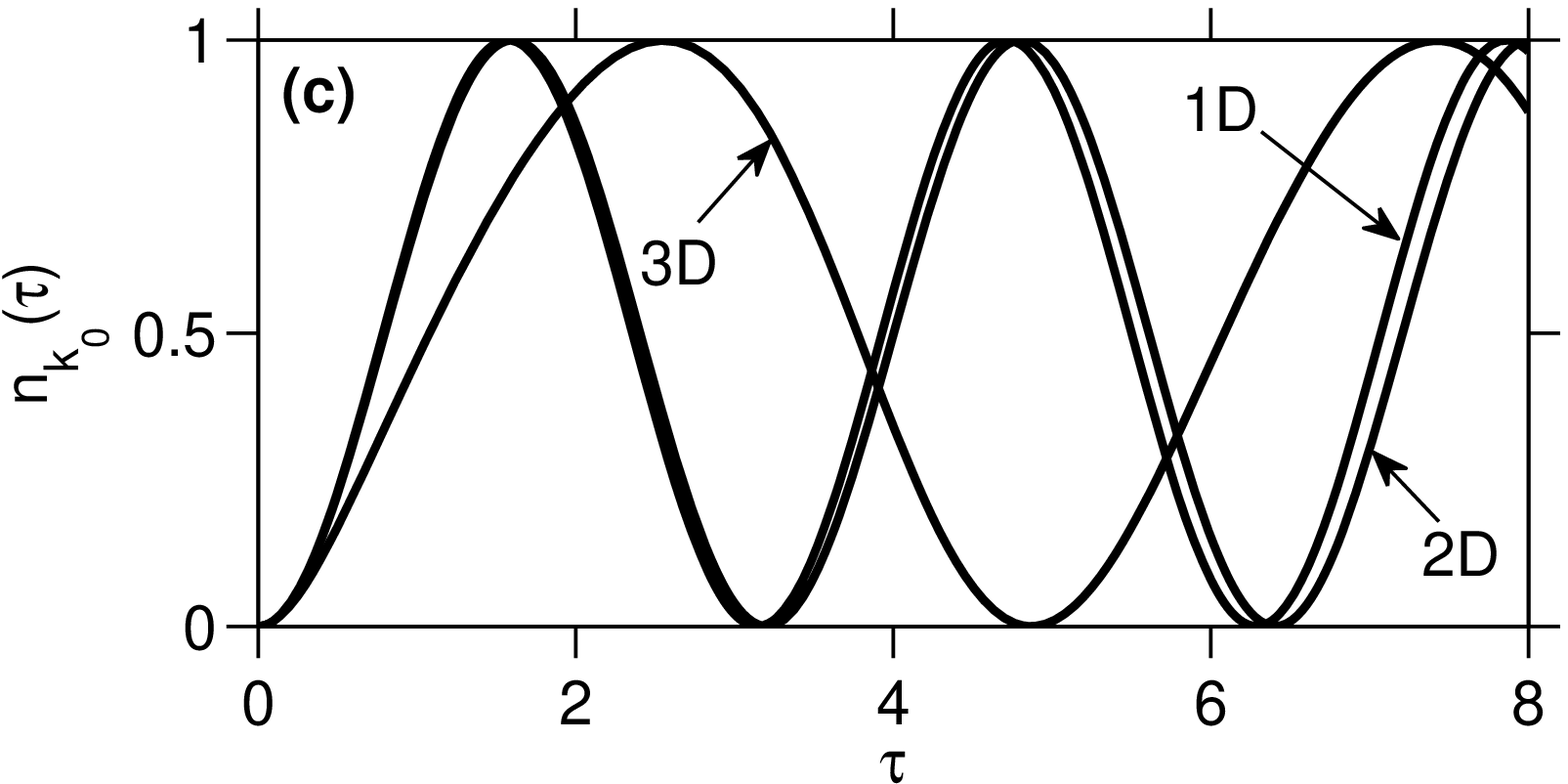}
\caption{Same as in Fig. \protect\ref{fig6} except for $|\protect\delta %
|=3174$. In (c) we show the temporal population of the resonant mode
$k_{0}$ in 1D, 2D, and 3D. The analytic result of $\sin
^{2}(\protect\tau )$ corresponding to the undepleted molecular field
approximation is practically identical to the 1D curve shown.}
\label{fig7}
\end{figure}
In this section we analyze the momentum distribution of the dissociated
atoms, $n_{\mathbf{k}}(t)$, which is the same for both spin states. In Fig.~%
\ref{fig6}(a) we plot the momentum distribution versus time for the case of
fermionic atom statistics. Due to the spherical symmetry in 3D, we only show
the radial dependence on the absolute momentum $k=|\mathbf{k}|$, scaled with
respect to the resonant momentum $k_{0}=\sqrt{2m_{1}|\Delta |/\hbar }$.

\begin{figure}[tbp]
\hspace{0.8cm}\includegraphics[height=3.95cm]{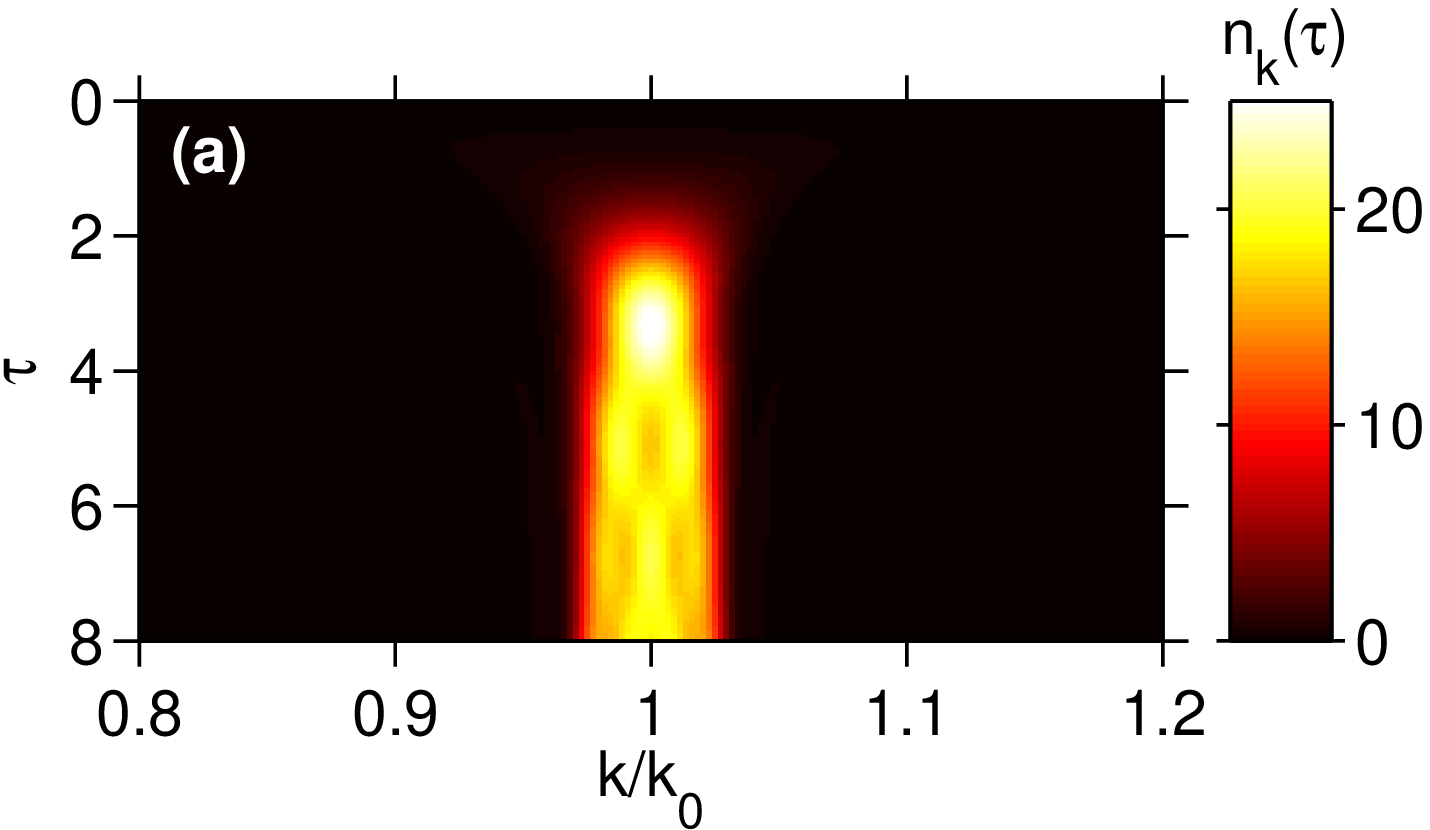}
\includegraphics[height=3.55cm]{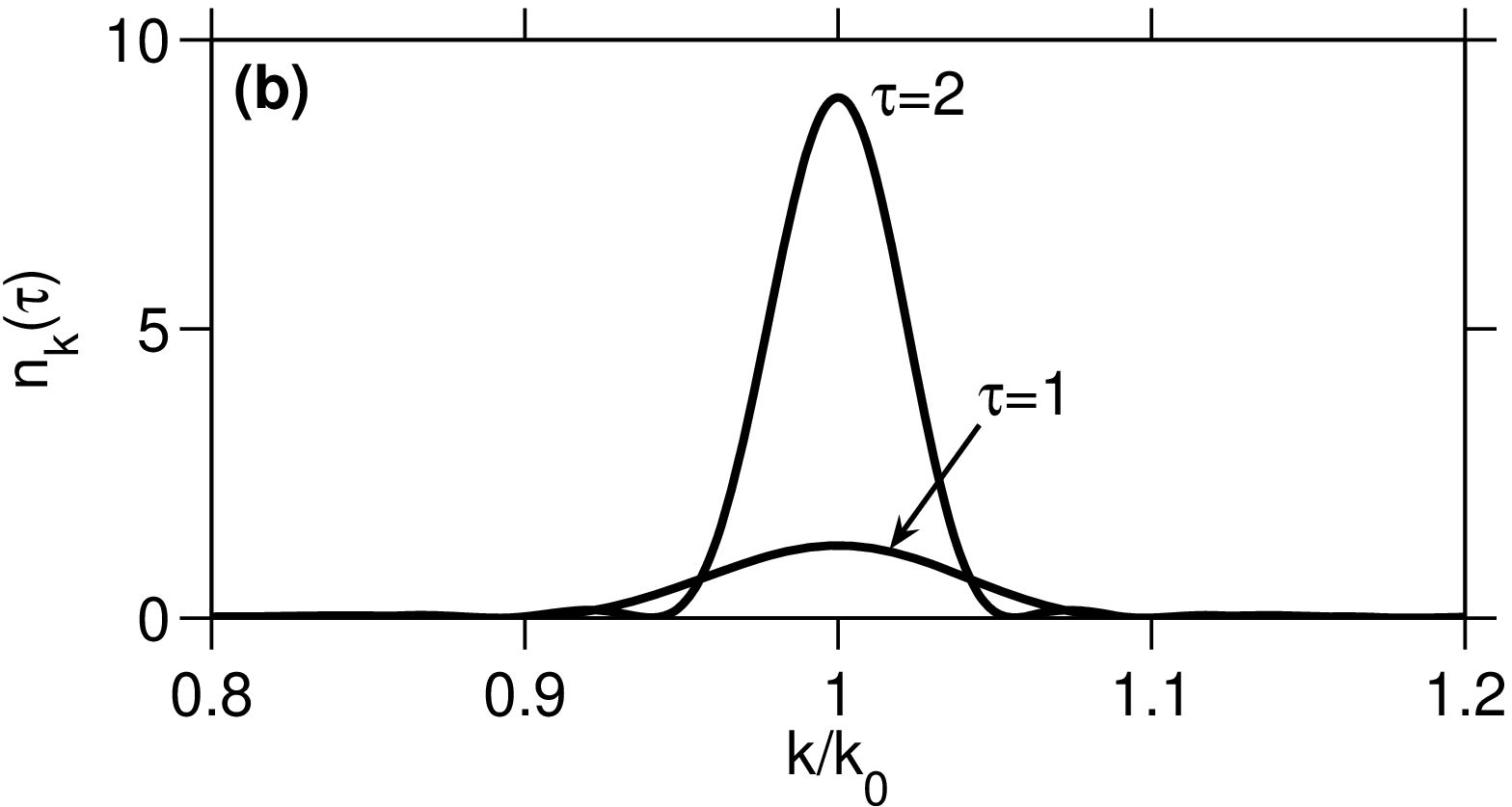}
\includegraphics[height=3.4cm]{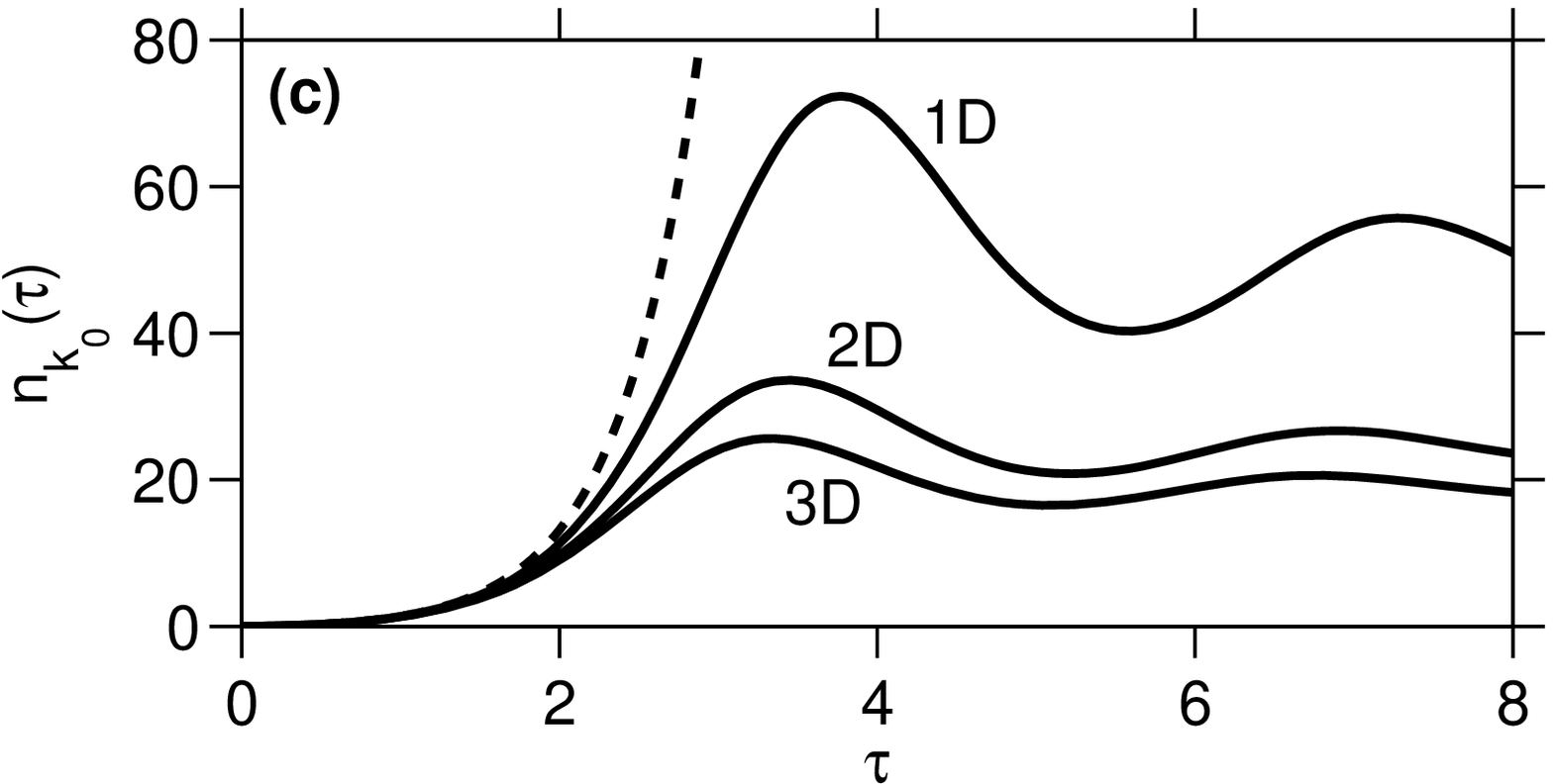}
\caption{Same as in Fig. \protect\ref{fig6} ($|\protect\delta |=16$) except
for bosonic statistics of the atoms. The time slices of the momentum
distribution in (b) are for $\protect\tau =1$ and $\protect\tau =2$. In (c),
we show the temporal population of the resonant mode $k_{0}$ in 1D, 2D, and 3D
(solid line), together with the respective analytic result, $\sinh ^{2}(%
\protect\tau )$ (dashed line), obtained using the undepleted molecular field
approximation \protect\cite{Fermidiss}.}
\label{fig8}
\end{figure}

\begin{figure}[tbp]
\hspace{0.8cm}\includegraphics[height=3.95cm]{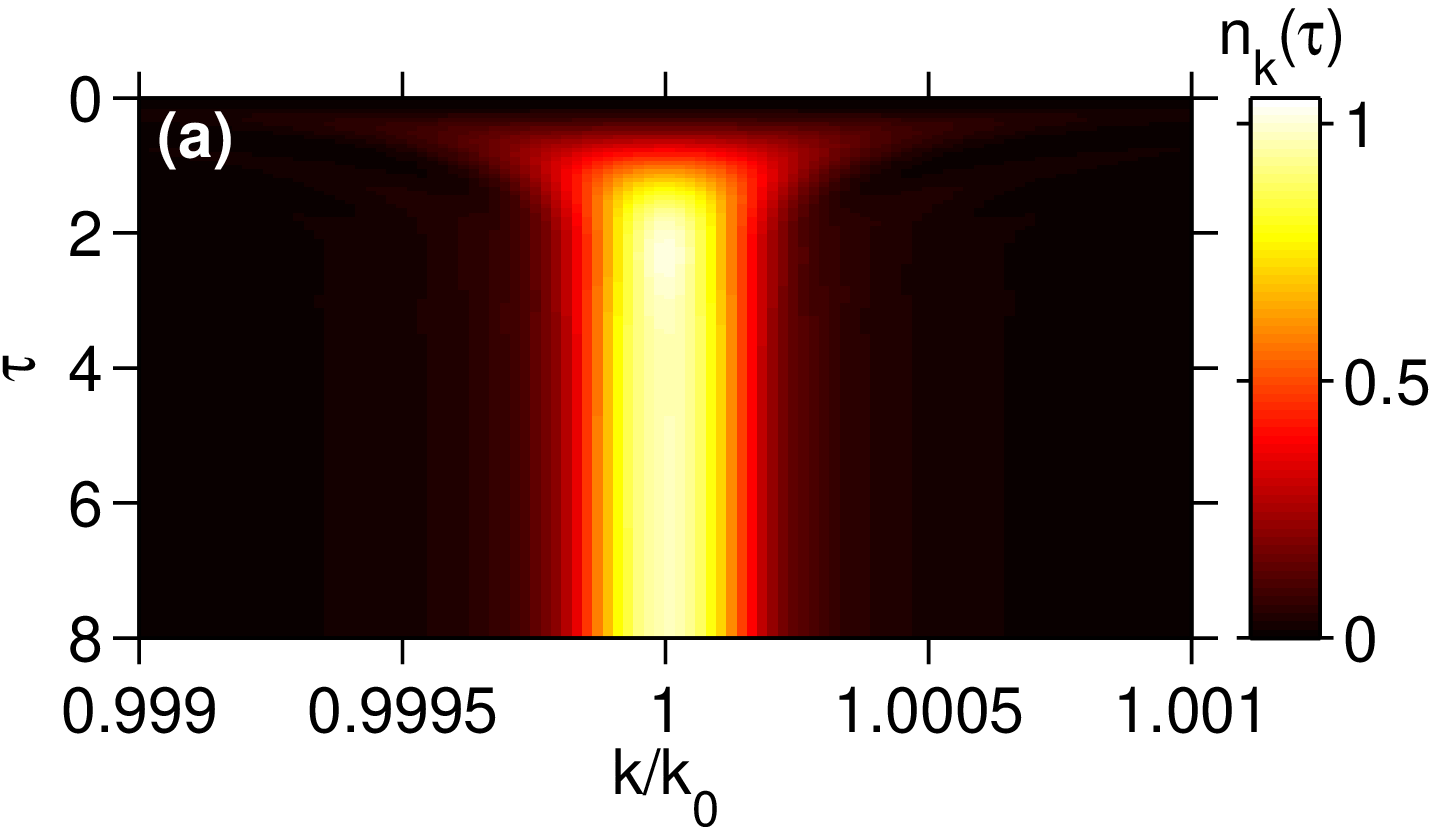}
\includegraphics[height=3.55cm]{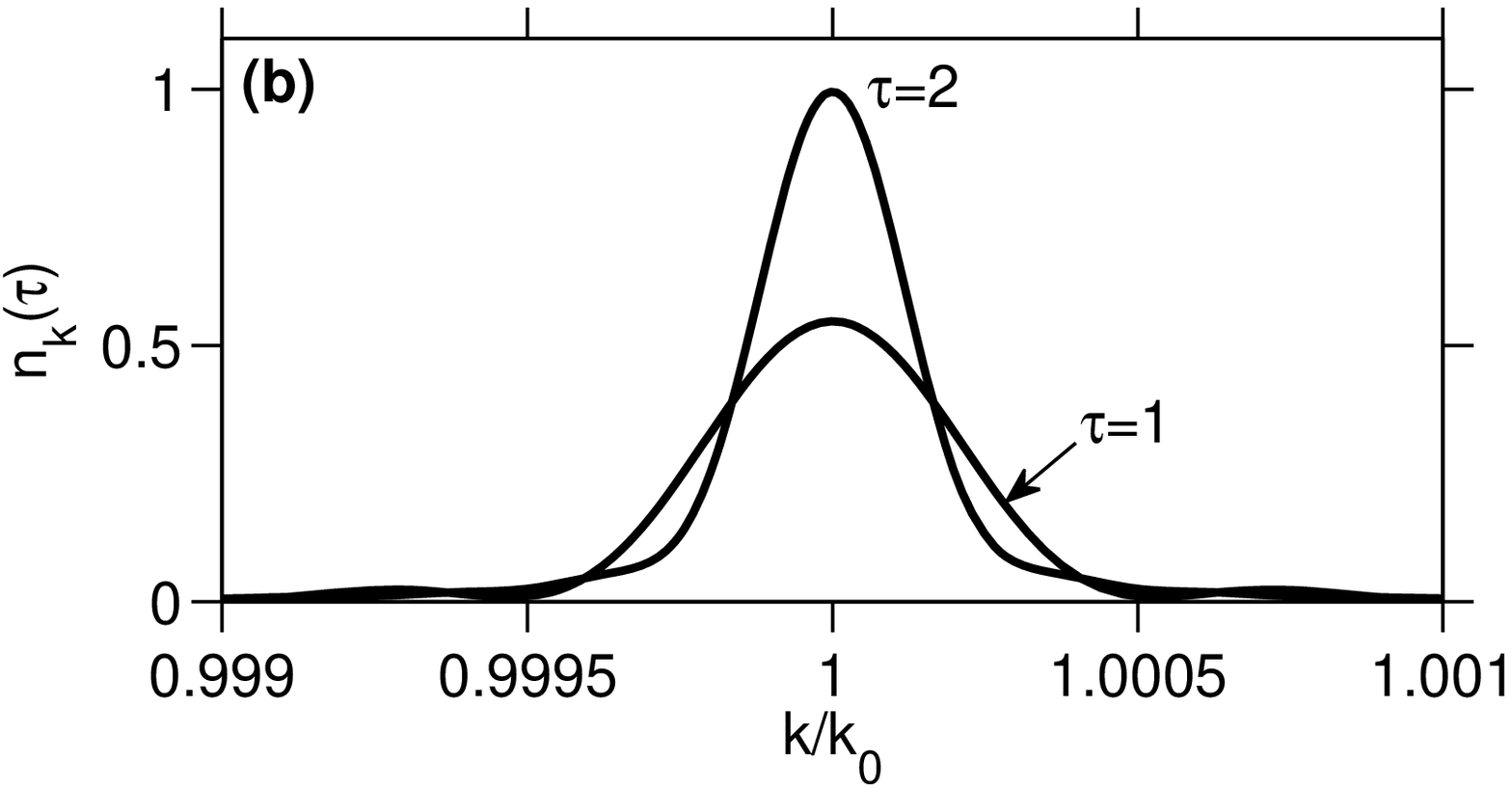}
\includegraphics[height=3.4cm]{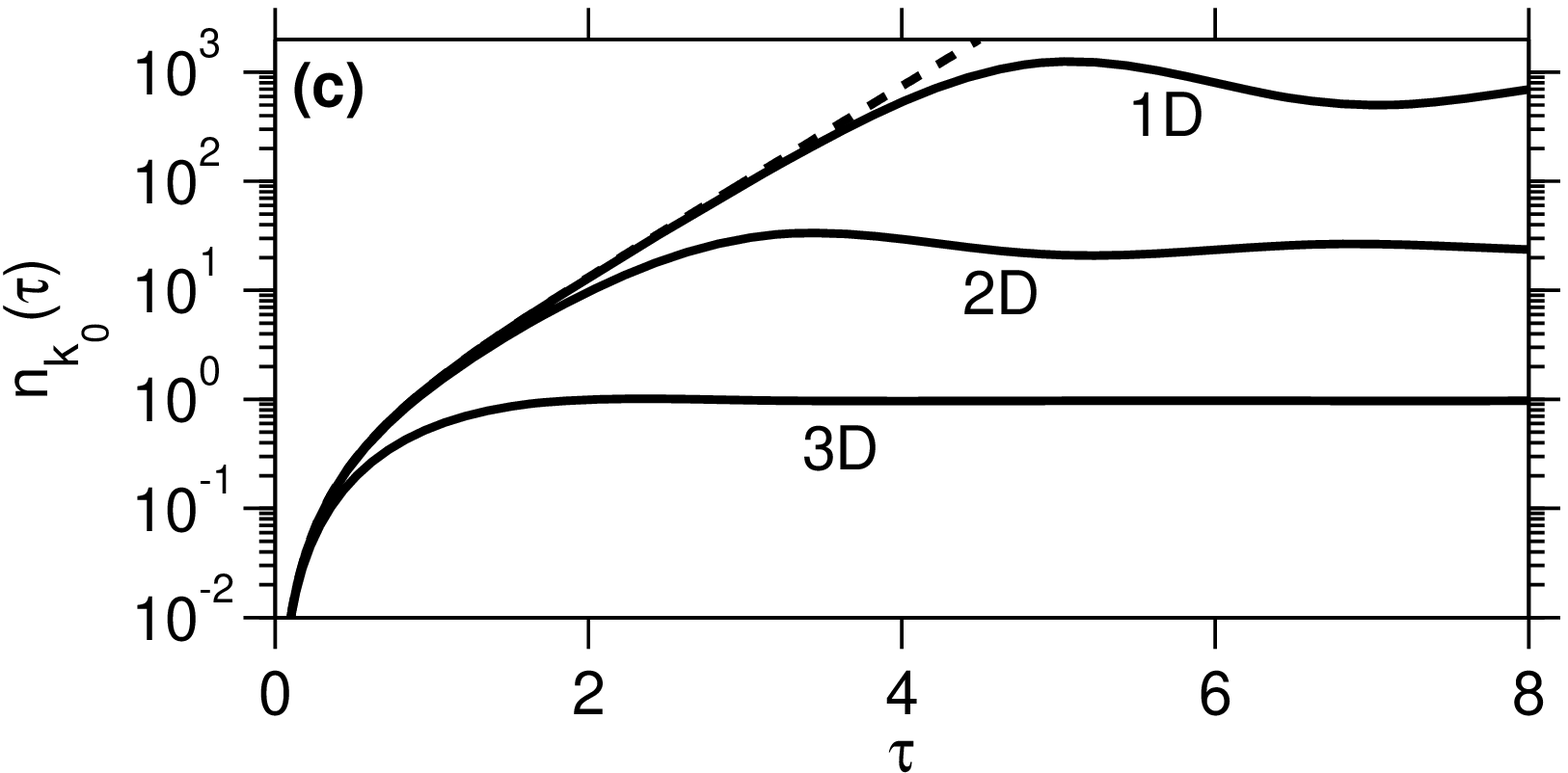}
\caption{Same as in Fig.~\protect\ref{fig7} ($|\protect\delta |=3174$)
except for bosonic atom statistics.}
\label{fig9}
\end{figure}

In Fig.~\ref{fig6}(b) we show examples of snapshots of the momentum
distribution at times $\tau =1$ and $\tau =3$. The respective 1D and
2D results are very similar to the 3D results and are not shown. The
maximum
deviation from the 3D result is in the 1D case and is less than $2\%$ at $%
\tau =3$.

In Fig.~\ref{fig6}(c) we plot the occupation of the resonant mode
$k_{0}$ as a function of time. The dashed line is the
analytic result obtained using the
undepleted molecular field approximation is given by $\sin^{2}(\tau )$, and
is applicable in all dimensions \cite{Fermidiss}.
Again the 1D and 2D results of the present pairing mean-field theory
are omitted for clarity --- in the 1D case the result is almost
indistinguishable from the dashed line, while in 2D it is
intermediate between the dashed line and the 3D solid line. The
larger discrepancy of the 3D result from the approximate analytic
solution is explained by the fact that molecular depletion is a
more significant effect in 3D. This is better seen in Fig.~\ref{fig7} for
the significantly larger absolute detuning $|\delta |$, where the
molecular depletion is more important and hence the discrepancy
between the pairing mean-field result and that of the undepleted
molecular field approximation is larger.

Figures \ref{fig8} and \ref{fig9} are the results for the same
parameters as in Figs.~\ref{fig6} and \ref{fig7} respectively except
that they are for bosonic atom statistics. Compared to the fermionic
case of Fig.~\ref{fig6}, the dissociation dynamics in
Figs.~\ref{fig8} are not affected by the Pauli exclusion
principle and we see larger discrepancies between the 1D, 2D and 3D
results. The disagreement between the undepleted molecular field
approximation and that of the present PMFT is again largest in the
3D case due to the larger density of states which results in faster
depletion.

It is instructive to use the above numerical results to determine
the width of the shell of the dissociation sphere and compare it
with the predictions of much simpler analytic approaches. By doing
this we can get a better understanding of the validity of these
approximate methods. For example, the simplest estimate of the
width of the dissociation sphere can be obtained from energy-time
uncertainty considerations. More specifically, for short duration of
dissociation $t$ ($\Gamma t\ll 1$) during which the dissociation is
in the spontaneous regime and is independent of atom statistics, the
width of the dissociation sphere can be obtained through the relation $%
\Delta Et\sim \hbar $. Here, the energy uncertainty is evaluated in terms of
the momentum uncertainty $\Delta k$ around the resonant momentum $k_{0}=%
\sqrt{2m_{1}|\Delta |/\hbar }$:%
\begin{equation}
\Delta E=\frac{\hbar ^{2}(k_{0}+\Delta k)^{2}}{2m_{1}}-\frac{\hbar
^{2}k_{0}{}^{2}}{2m_{1}}\simeq \frac{\hbar ^{2}k_{0}\Delta k}{m_{1}},
\end{equation}%
where we assumed $\Delta k/k_{0}\ll 1$. According to this, the momentum
uncertainty $\Delta k$ which gives the width of the dissociation sphere is
given by
\begin{equation}
\frac{\Delta k}{k_{0}}\simeq \frac{m}{\hbar k_{0}^{2}t}=\frac{1}{2|\Delta |t}%
=\frac{1}{2|\delta |\tau },  \label{dleta-k}
\end{equation}%
which is inversely proportional to the duration of dissociation $t$
as expected and also to $k_{0}^{2}$ (or $|\Delta|$), implying that
for larger detuning $|\Delta |$ the width of the sphere is narrower.
These conclusions are clearly seen from our numerical results for
$\tau \lesssim 1$, shown in Figs.~\ref{fig6}--\ref{fig9}. For
example, at $\tau =1$ we obtain [from Eq.~(\ref{dleta-k})]
$\Delta k/k_{0}\simeq 0.03$ for $|\delta |=16$ and $\Delta
k/k_{0}\simeq 1.6\times 10^{-4}$ for $|\delta |=3174$, in good
agreement with the numerical results of
Figs.~\ref{fig6}(b)--\ref{fig9}(b).

For longer duration of dissociation, when the dynamics is strongly
affected by either Bose stimulation or Pauli blocking, the width of
the dissociation sphere can be estimated (see Appendix
\ref{sect:appendix-D}) using a slightly more involved approximate
approach based on analytic solutions within the undepleted molecular
BEC approximation \cite{Fermidiss,Savage}. In the fermionic case,
where the momentum spectrum develops oscillatory peaks, the width in
question is for the envelope function. Apart from this distinction
between the fermionic and bosonic results, the width of the
dissociation spherical shell is the same in the two cases and is
given by
\begin{equation}
\frac{\Delta k}{k_{0}}\simeq \frac{m\chi _{D}\sqrt{N_{0}/L^{D}}}{2\hbar
k_{0}^{2}}=\frac{1}{4|\delta |}.  \label{width-stim}
\end{equation}%
This is still inversely proportional to the dimensionless detuning
$|\delta | $, but no longer depends on time as the momentum
distribution in this regime ($\tau \gtrsim 2$ in most of our
examples) settles into a shape with an approximately constant
(envelope) width and residual oscillations [see
Figs.~\ref{fig6}--\ref{fig9}]. For $|\delta |=16$ and $|\delta
|=3174$, the above expression gives, respectively, $\Delta
k/k_{0}=0.016$ and $\Delta k/k_{0}=0.79\times 10^{-4}$ which are
again in good agreement with the numerical results.

\subsection{Atom-atom correlations}
\label{aa_corr}
\subsubsection{Distinguishable case}

In this section we study the correlation between the occupation number
fluctuations for atoms in opposite spin states
\begin{eqnarray}
G_{12}(\mathbf{k},\mathbf{k}^{\prime },t) &=&\frac{\langle \Delta \hat{n}_{1,%
\mathbf{k}}(t)\Delta \hat{n}_{2,\mathbf{k}^{\prime }}(t)\rangle }{\langle
\hat{n}_{1,\mathbf{k}}(t)\rangle ^{1/2}\langle \hat{n}_{2,\mathbf{k}^{\prime
}}(t)\rangle ^{1/2}} \notag \\
&=&\frac{\langle \hat{n}_{1,\mathbf{k}}(t)\hat{n}_{2,\mathbf{k}^{\prime
}}(t)\rangle -\langle \hat{n}_{1,\mathbf{k}}(t)\rangle \langle \hat{n}_{2,%
\mathbf{k}^{\prime }}(t)\rangle }{\langle \hat{n}_{1,\mathbf{k}}(t)\rangle
^{1/2}\langle \hat{n}_{2,\mathbf{k}^{\prime }}(t)\rangle ^{1/2}},  \notag \\
&&\;  \label{G12-def}
\end{eqnarray}%
where $\Delta \hat{n}_{j,\mathbf{k}}(t)=\hat{n}_{j,\mathbf{k}}(t)-\langle
\hat{n}_{j,\mathbf{k}}(t)\rangle $ is the fluctuation. The definition of $%
G_{12}$ is different from the more conventional Glauber second-order
correlation function \cite{Glauber}%
\begin{equation}
g_{12}^{(2)}(\mathbf{k},\mathbf{k}^{\prime },t)=\frac{\langle \hat{a}_{1,%
\mathbf{k}}^{\dagger }(t)\hat{a}_{2,\mathbf{k}^{\prime }}^{\dagger }(t)\hat{a%
}_{2,\mathbf{k}^{\prime }}(t)\hat{a}_{2,\mathbf{k}}(t)\rangle }{\langle \hat{%
n}_{2,\mathbf{k}}(t)\rangle \;\langle \hat{n}_{2,\mathbf{k}^{\prime
}}(t)\rangle },
\end{equation}%
in that $G_{12}$ is normalized to the square root of the product of the
occupation numbers (rather than to the direct product) and the contribution
from uncorrelated statistics is subtracted from it so that absence of
correlation corresponds to $G_{12}=0$. The above definition of $G_{12}$ has
been used in Ref.~\cite{Greiner} for quantifying the strength of correlation
between the atoms in two hyperfine states of $^{40}$K atoms created through
molecular dissociation, and it is particularly convenient for fermionic atom
statistics since in this case $G_{12}$ is bounded between $0$ and $1$. The
correlation functions $G_{12}$ and $g_{12}^{(2)}$ are related by a simple
relationship,

\begin{equation}
g_{12}^{(2)}(\mathbf{k},\mathbf{k}^{\prime },t)=1+\frac{G_{12}(\mathbf{k},%
\mathbf{k}^{\prime },t)}{\langle \hat{n}_{1,\mathbf{k}}(t)\rangle
^{1/2}\langle \hat{n}_{2,\mathbf{k}^{\prime }}(t)\rangle ^{1/2}}.
\end{equation}

Using Wick's theorem to factore higher-order correlation functions and
 express them in
terms of the normal and anomalous populations, $\langle \hat{a}_{i,\mathbf{k}%
}^{\dagger }(t)\hat{a}_{j,\mathbf{k}^{\prime }}(t)\rangle $ and $\langle
\hat{a}_{i,\mathbf{k}}(t)\hat{a}_{j,\mathbf{k}^{\prime }}(t)\rangle $, we
find, in particular, that the correlation between the atoms with equal but
opposite momenta, $\mathbf{k}^{\prime }=-\mathbf{k}$, is given by%
\begin{equation}
G_{12}(\mathbf{k},-\mathbf{k},t)=\frac{|m_{\mathbf{k}}(t)|^{2}}{n_{\mathbf{k}%
}(t)}=1\pm n_{\mathbf{k}}(t).  \label{G12}
\end{equation}%
Here, the upper (lower) sign refers to bosonic (fermionic)
statistics, and we have used the fact that in the present
pairing-mean field theory any second-order moments other than those
in Eqs.~(\ref{nk-def}) and (\ref{mk-def}) are zero. In addition, we
have used Eq.~(\ref{eq:nm}) for expressing the anomalous occupation
$m_{\mathbf{k}}(t)$ via the normal occupation number
$n_{\mathbf{k}}(t)$. This gives the maximum possible value of the
anomalous moment $m_{\mathbf{k}}$ and hence the maximum strength of
the second-order correlation function between the atoms with
opposite momenta \cite{Fermidiss,Savage}. Therefore the results of
the present (approximate) PMFT for correlation functions give the
respective upper bounds to their exact values.

For any $\mathbf{k}^{\prime }\neq -\mathbf{k}$,
on the other hand, we have%
\begin{equation}
G_{12}(\mathbf{k},\mathbf{k}^{\prime },t)=0,\;\;\mathbf{k}^{\prime }\neq -%
\mathbf{k}  \label{G12-0}
\end{equation}%
which corresponds to absence of any correlation.

For completeness, we also give the results for the correlation between the
occupation fluctuations for the atoms in the same spin state:%
\begin{equation}
G_{jj}(\mathbf{k},\mathbf{k}^{\prime },t)=\frac{\langle \Delta \hat{n}_{j,%
\mathbf{k}}(t)\Delta \hat{n}_{j,\mathbf{k}^{\prime }}(t)\rangle }{\langle
\hat{n}_{j,\mathbf{k}}(t)\rangle ^{1/2}\langle \hat{n}_{j,\mathbf{k}^{\prime
}}(t)\rangle ^{1/2}}.
\end{equation}%
For the present pairing mean-field theory, this is given by%
\begin{equation}
G_{11}(\mathbf{k},\mathbf{k}^{\prime },t)=G_{22}(\mathbf{k},\mathbf{k}%
^{\prime },t)=\left\{
\begin{array}{l}
\pm n_{\mathbf{k}}(t),\;\;\mathbf{k}^{\prime }=\mathbf{k,} \\
0,\;\;\;\;\mathbf{k}^{\prime }\neq \mathbf{k},%
\end{array}%
\right.
\end{equation}%
where again the upper (lower) sign refers to bosonic (fermionic) statistics.
Here, the case with $\mathbf{k}^{\prime }\neq \mathbf{k}$ implies
uncorrelated statistics, which in terms of Glauber's second-order
correlation function corresponds to $g_{11}^{(2)}(\mathbf{k},\mathbf{k}
^{\prime }\neq \mathbf{k},t)=1$. Similarly, $G_{11}(\mathbf{k},\mathbf{k}
,t)=n_{\mathbf{k}}(t)$ in the bosonic case corresponds to the level of
correlations between thermally bunched atoms, $g_{11}^{(2)}(\mathbf{k},
\mathbf{k},t)=2$. Finally, $G_{11}(\mathbf{k},\mathbf{k},t)=-n_{\mathbf{k}
}(t)$ in the fermionic case corresponds to $g_{11}^{(2)}(\mathbf{k},\mathbf{%
k },t)=0$ implying perfect antibunching, which is a simple consequence of
the Pauli exclusion principle.

The result of Eq.~(\ref{G12}) for the opposite spin atoms corresponds to
\begin{equation}
g_{12}^{(2)}(\mathbf{k},-\mathbf{k},t)=1+\frac{1\pm n_{\mathbf{k}}(t)}{n_{%
\mathbf{k}}(t)}.
\end{equation}%
Thus, strong correlation between the atoms with equal but opposite momenta
corresponds to $g_{12}^{(2)}(\mathbf{k},-\mathbf{k},t)=1/n_{\mathbf{k}}(t)>1$
in the fermionic case, with $n_{\mathbf{k}}(t)<1$ due to Pauli blocking, and
to $g_{12}^{(2)}(\mathbf{k},-\mathbf{k},t)=2+1/n_{\mathbf{k}}(t)>2$ in the
bosonic case, corresponding to super-thermal bunching.

Since Wick's factorization scheme as employed here is equally valid for the
pairing mean-field theory and for the treatment of Ref.~\cite{Fermidiss}
with the undepleted molecular field approximation, all results for
second-order correlation functions obtained in Ref.~\cite{Fermidiss} remain
valid here. The only difference is that the analytic solutions for $n_{%
\mathbf{k}}(t)$ \cite{Fermidiss} are now replaced by the numerical solution
of Eqs.~(\ref{PMFT}). This takes into account molecular depletion and
therefore the present results are expected to be more accurate.

\subsubsection{Indistinguishable case}

In the case of molecular dissociation into bosonic atom pairs in the same
spin state, the considerations and analysis of the previous subsection
remain valid, except that the role of $G_{12}(\mathbf{k},-\mathbf{k},t)$ is
now taken by $G_{11}(\mathbf{k},-\mathbf{k},t)$. We note, however, that for
indistinguishable atoms the conveniences associated with using $G_{11}(%
\mathbf{k},\mathbf{k}^{\prime },t)$ -- defined analogously to Eq. (\ref%
{G12-def}) -- for characterizing the strength of correlation between atom
pairs are no longer valid. For example, in the present pairing mean-field
theory we find that the correlation functions $G_{11}(\mathbf{k},-\mathbf{k}%
,t)$ and $G_{11}(\mathbf{k},\mathbf{k},t)$ are both equal to $1+n_{\mathbf{k}%
}(t)$, which does not distinguish between the strength of correlation
for atom pairs with opposite momenta and the strength of autocorrelation
signal. For this reason, we use Glauber's second-order correlation function
\begin{equation}
g_{11}^{(2)}(\mathbf{k},\mathbf{k^{\prime }},t)=\frac{\langle \hat{a}_{1,%
\mathbf{k}}^{\dagger }(t)\hat{a}_{1,\mathbf{k}^{\prime }}^{\dagger }(t)\hat{a%
}_{1,\mathbf{k}^{\prime }}(t)\hat{a}_{1,\mathbf{k}}(t)\rangle }{\langle \hat{%
n}_{1,\mathbf{k}}(t)\rangle \;\langle \hat{n}_{1,\mathbf{k}^{\prime
}}(t)\rangle }.  \label{g2}
\end{equation}

Following the same arguments as in Ref. \cite{Savage}, we obtain the
following results
\begin{equation}
g_{11}^{(2)}(\mathbf{k},\mathbf{k}^{\prime },t)=\left\{
\begin{array}{l}
1,\,\;\quad \quad \quad \quad \,\,\mathbf{k}^{\prime }\neq \mathbf{k},%
\mathbf{k}^{\prime }\neq -\mathbf{k}, \\
2,\,\;\quad \quad \quad \quad \,\,\mathbf{k}^{\prime }=\mathbf{k}\neq 0, \\
2+1/n_{\mathbf{k}}(t),\,\,\mathbf{k}^{\prime }=-\mathbf{k}, \\
3+1/n_{\mathbf{k}}(t),\,\,\mathbf{k}^{\prime }=\mathbf{k}=0,%
\end{array}%
\right.   \label{analytic correlation functions}
\end{equation}%
which again formally coincide with those obtained using the
undepleted molecular field approximation~\cite{Savage}, except that
the mode occupation numbers $n_{\mathbf{k}}(t)$ are now given by the
numerical solutions of the present pairing mean-field theory,
Eqs.~(\ref{PMFT}) and (\ref{PMFT-samespin}).

\subsection{Correlations in column densities in 3D systems}

\begin{figure}[tbp]
\includegraphics[height=3.5cm]{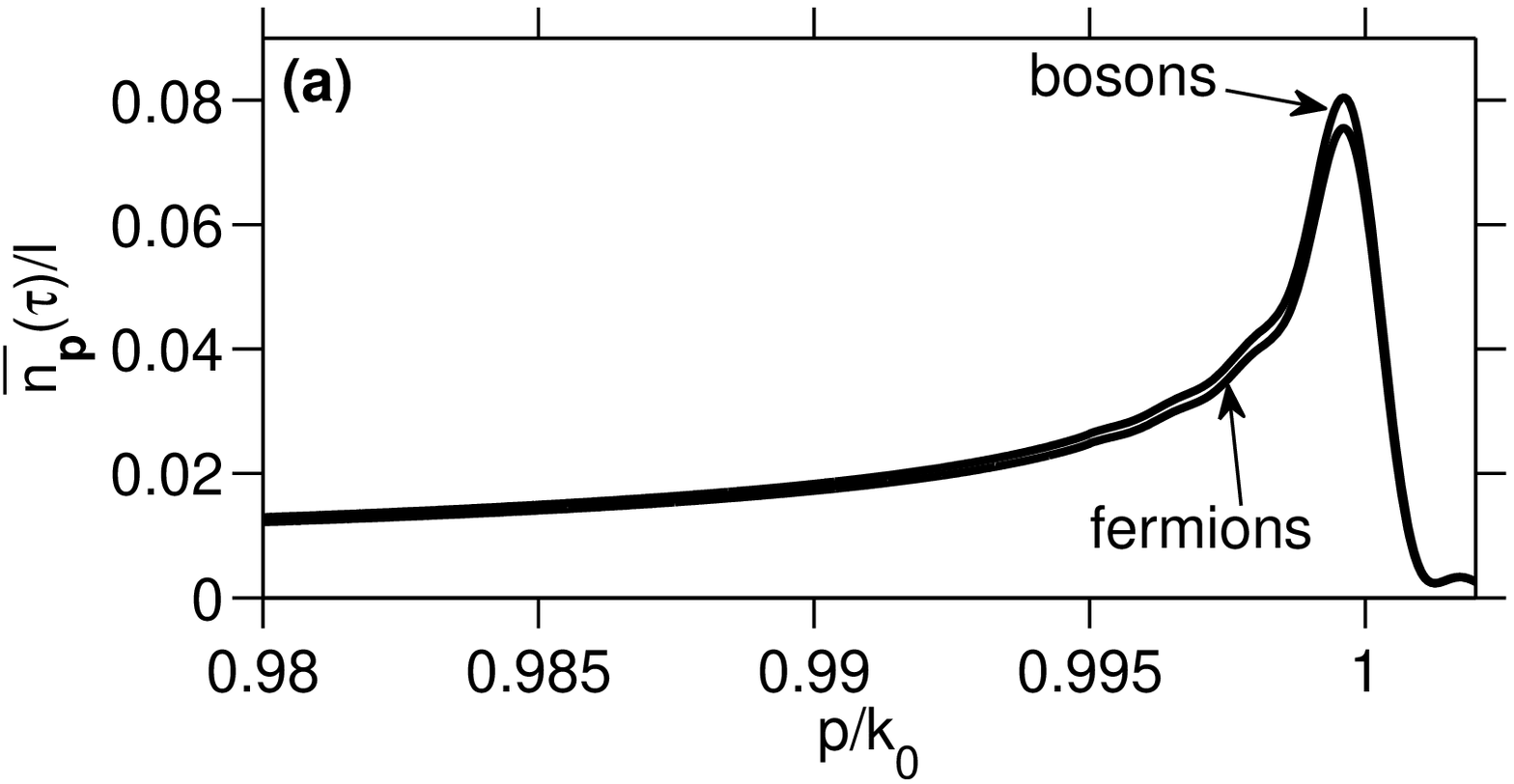}
\includegraphics[height=3.5cm]{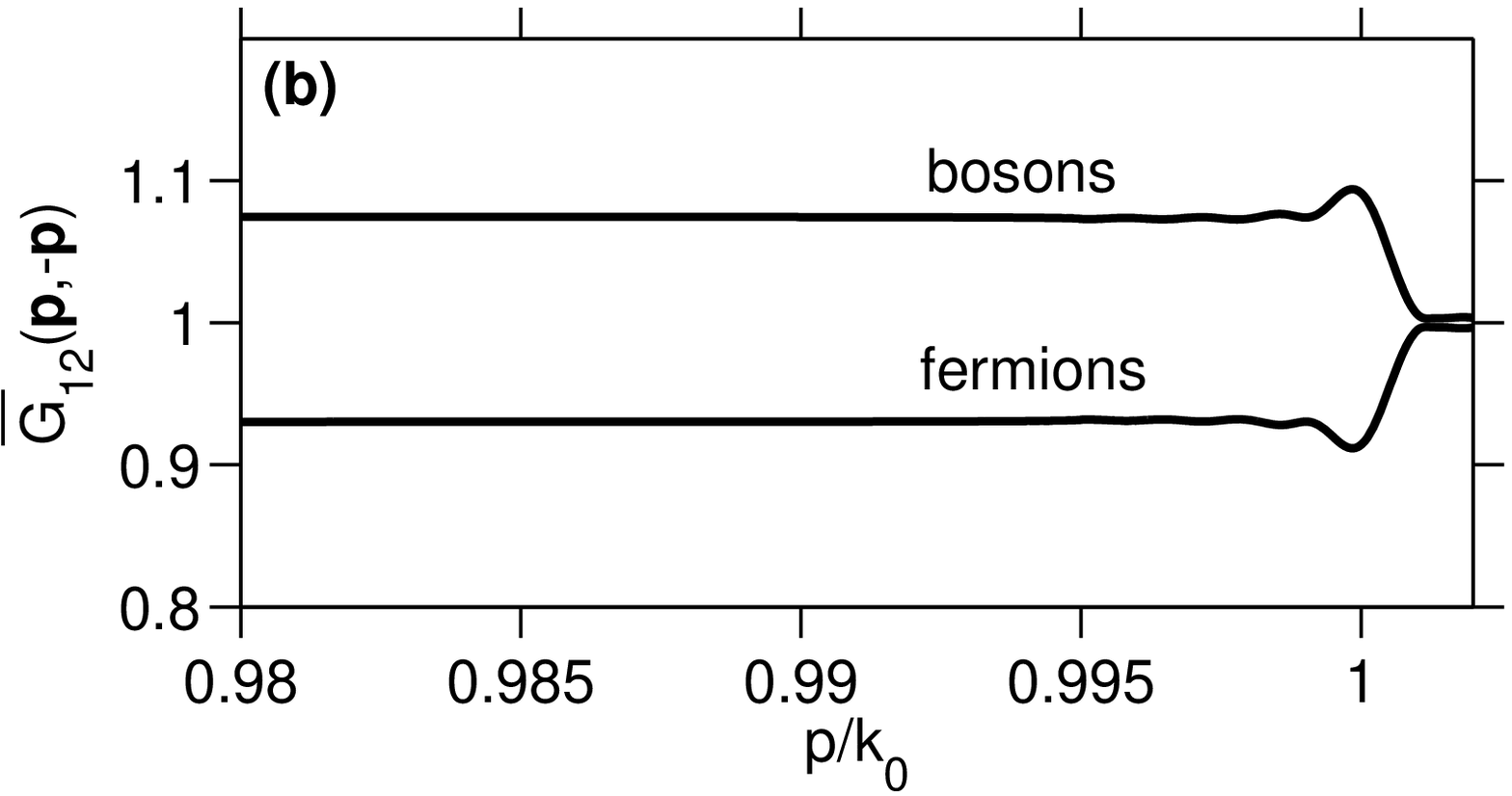}
\caption{(a) Momentum-space column density $\overline{n}_{\mathbf{p}}(%
\protect\tau )/l$ of each spin component at $\protect\tau =0.36$ for bosonic
and fermionic cases in 3D for $|\protect\delta |=3174$ and $%
N_{0}(0)/l^{D}=3.1$. The horizontal axis is the absolute value of the 2D
reduced momentum $p=|\mathbf{p}|$, where $\mathbf{p\equiv }(k_{x},k_{y})$.
(b) Correlation function between momentum column density fluctuations for
the atoms with opposite spins and conjugate momenta, $\overline{G}_{12}(%
\mathbf{p},-\mathbf{p},\protect\tau )$, as a function of $p=|\mathbf{p}|$,
for the same parameter values as in (a). }
\label{fig10}
\end{figure}

The correlation functions derived and discussed in the previous section are
difficult to measure experimentally. Rather than density functions such as $%
n(\mathbf{x})$ being measured in the laboratory, column densities such as $%
\int dz n(\mathbf{x})$ are obtained using the technique of absorption
imaging. Indeed, the correlation measurements performed in the experiments
at JILA on dissociation of a molecular condensate of $^{40}$K$_{2}$ dimers
into fermionic atom pairs~\cite%
{Greiner} were made using absorption images after a time-of-flight
expansion. We now analyze the momentum space analog of this procedure and
calculate the correlation between momentum column density fluctuations.

First, we define the atom number operator corresponding to a $z$-integrated
momentum column density, $\widehat{\overline{n}}_{j,\mathbf{p}%
}=\sum\nolimits_{k_{z}}\hat{n}_{j,\mathbf{k}}$, where $\mathbf{p\equiv }%
(k_{x},k_{y})$ is the reduced 2D momentum. The average column density $%
\overline{n}_{\mathbf{p}}(t)$ $=\langle \widehat{\overline{n}}_{j,\mathbf{p}%
}(t)\rangle $ is found via $\overline{n}_{\mathbf{p}}(t)=\sum%
\nolimits_{k_{z}}n_{\mathbf{k}}(t)$ and is the same for the two spin states,
$j=1,2$. The correlation function between the momentum column density
fluctuations in the two spin states, which we denote via $\overline{G}_{12}(%
\mathbf{k},\mathbf{k}^{\prime },t)$, is defined as in Eq. (\ref{G12-def})
except that the operators $\hat{n}_{j,\mathbf{k}}$ are replaced by $\widehat{%
\overline{n}}_{j,\mathbf{p}}$:%
\begin{equation}
\overline{G}_{12}(\mathbf{p},\mathbf{p}^{\prime },t)=\frac{\langle \Delta
\widehat{\overline{n}}_{1,\mathbf{p}}(t)\Delta \widehat{\overline{n}}_{2,%
\mathbf{p}^{\prime }}(t)\rangle }{\langle \widehat{\overline{n}}_{1,\mathbf{p%
}}(t)\rangle ^{1/2}\langle \widehat{\overline{n}}_{2,\mathbf{p}^{\prime
}}(t)\rangle ^{1/2}}.
\end{equation}%
We emphasize that the summation over the $q_{z}$-component has to be
performed before taking the ensemble average.

Using Wick's theorem as before, we obtain the following results for equal
but opposite momenta, $\mathbf{p}^{\prime }=-\mathbf{p}$:
\begin{equation}
\overline{G}_{12}(\mathbf{p},-\mathbf{p},t)=1\pm \frac{\sum\nolimits_{k_{z}}%
\left[ n_{\mathbf{k}}(t)\right] ^{2}}{\overline{n}_{\mathbf{p}}(t)}.
\label{corr-CD}
\end{equation}%
Here the plus sign stands for bosons, while the minus sign is for fermions,
and the distinction between the bosonic and fermionic results is highlighted
through the fact that $\overline{G}_{12}(\mathbf{p},-\mathbf{p},t)>1$\ for
bosons and $\overline{G}_{12}(\mathbf{p},-\mathbf{p},t)<1$ for fermions. For
any other pair of momenta ($\mathbf{p}^{\prime }\neq -\mathbf{p}$), the
correlation function is zero, implying the absence of any correlation.

In Fig.~\ref{fig10} we show examples of momentum column densities $\overline{%
n}_{\mathbf{p}}(t)$ and correlations between column density fluctuations $%
\overline{G}_{12}(\mathbf{p},-\mathbf{p},t)$ for bosonic and
fermionic atoms for $|\delta |=3174$ at $\tau =0.36$. The
dissociation at this stage is still in the spontaneous regime, so
the difference between the column densities for bosonic and
fermionic cases is hardly noticeable in Fig.~\ref{fig10}(a). On the
other hand, the difference due to quantum statistics becomes much
more pronounced if one examines the pair correlation function
$\overline{G}_{12}(\mathbf{p},-\mathbf{p},t)$, as shown
Fig.~\ref{fig10}(b), with $\overline{G}_{12}(\mathbf{p},-%
\mathbf{p},t)>1$ for bosons and $\overline{G}_{12}(\mathbf{p},-\mathbf{p}%
,t)<1$ for fermions.

The example of Fig.~\ref{fig10} in the fermionic case, with $\overline{G}%
_{12}(\mathbf{p},-\mathbf{p},t)\simeq 0.9$, corresponds to the
experimental parameters of Ref. \cite{Greiner}. We note that the
experimentally measured spatial counterpart of the same correlation
function after time-of-flight expansion varied between $0.1$
and $0.3$ depending on the bin size used in the analysis of the
imaged data. The dependence of the strength of the correlation
signal on bin size was analyzed theoretically in Ref.
\cite{SavageKheruntsyanSpatial}; However,
ignoring variations due to binning, the experimental result was significantly less
than
our theoretically calculated value in momentum space.  This highlights
an important limitation of the present uniform theory, which is
that it only gives the \textit{upper bound} of correlation functions \cite%
{Fermidiss} and is \textit{not} adequate for correct quantitative
description of realistic nonuniform systems. This conclusion,
however, does not necessarily apply to other quantities of interest,
such as the momentum distribution or the total atom number. In these
cases, the quantitative predictions of the uniform theory can
adequately describe nonuniform systems if the uniform system is
properly \textit{size-matched}
to the nonuniform one as done in Sec.~\ref{sect:total-N} (see also Ref.~\cite%
{Savage}).

The reason that the PMFT gives unsatisfactory results for pair
correlations is the mode-mixing which couples the atomic momenta not
only in a pair-wise fashion, ($\mathbf{k}$,$-\mathbf{k}$), but also
couples within the neighborhood of the
partners. As a result the correlations are broadened, while their
strength can be dramatically reduced, especially in systems with
strong inhomogeneity. The problem of mode-mixing was analyzed in
detail in the \textit{bosonic} case in Ref.~\cite{Savage} using the
comparison between uniform solutions and exact quantum simulations
of nonuniform systems using the stochastic positive
$P$-representation method. Performing similar first-principle
quantum dynamical calculations for nonuniform, multimode
\textit{fermionic} systems is a challenging problem yet to be
solved.

\section{Conclusions}

\label{sect:Conclusions}

We have developed a simple pairing mean-field theory for the quantum
dynamics of the dissociation of a Bose-Einstein condensate of
diatomic molecules into its constituent particles, covering the
cases of indistinguishable bosons and distinguishable fermions or
bosons. The pairing fields for the atoms are introduced at the level
of normal and anomalous moments for the atomic creation and
annihilation operators; such an approach is necessary as the
dissociation is initially a spontaneous process and cannot be
described by ordinary mean field theory. We have solved the
resulting equations of motion for all cases, and identified the role
of dimensionality and quantum statistics on the conversion rate, and
compared them to the simpler analytic solutions available within the
undepleted molecular field approximation.

We find that molecular depletion is a more important effect in higher
dimensions and in the case of dissociation into bosonic atoms.
Accordingly, we argue that the undepleted molecular approximation is
more reliable for describing dissociation into fermionic atoms and
systems of reduced dimensionality. This conclusion can be useful in
extensions of the theory of dissociation to spatially inhomogeneous
systems. In this case, the simplest treatment would be to employ the
undepleted molecular field approximation, while still being able to
address the problem of mode-mixing, which is the most important one
when seeking quantitative description of atom-atom correlations.

We have also compared the results in the bosonic, indistinguishable case to the
exact dynamics found by positive-$P$ simulation and identified the
range of validity of the pairing mean-field theory. In addition, we
have compared our formalism to that of Jack and Pu \cite{Jack-Pu},
and identified the key differences. Finally, we have derived and
calculated results for the non-trivial correlations that are present
in the dissociated atoms, and compared our results to those of the
experiment performed at JILA~\cite{Greiner}.

The main limitation of this work (see also Sec.
\ref{sect:approximations}) is the assumption of a uniform molecular
condensate. This limits the time for which the dynamics are valid
and also does not allow for the association of atoms into previously
unoccupied molecular modes. The PMFT can be used to describe
realistic nonuniform condensates provided the results are applied to
a size-matched system. The total atom number and atomic density
distribution obtained in this way can be quantitatively reasonable,
however, the same is not true if the calculated quantities involve
density-density correlation functions. In this case the PMFT
provides upper bounds for the correlation functions. We are
currently extending the PMFT to include the effects of spatial
inhomogeneity to provide a quantitatively better model.

\section*{Acknowledgements}

The authors acknowledge support of this work by the Australian Research
Council and the Queensland State Government.

\appendix

\section{Decay rate in the distinguishable case}

\label{sect:appendix-A}

The transition matrix element $V_{ma}$ in Eq. (\ref{Gamma-def}) is
\begin{equation}
V_{ma}=\langle f|\hat{H}_{int}|i\rangle ,  \label{Vma-def}
\end{equation}%
where $\hat{H}_{int}$ is the atom-molecule coupling term in the Hamiltonian (%
\ref{eq:Ham}), while the initial $|i\rangle $ and final $|f\rangle $ states
describe a free molecule or a pair of free atoms respectively
\begin{eqnarray}
|i\rangle &=&\frac{1}{L^{D/2}}\int d^{D}\mathbf{x}\hat{\Psi}_{0}^{\dagger }(%
\mathbf{x})e^{i\mathbf{K\cdot x}}|0\rangle ,  \label{i} \\
|f\rangle &=&\frac{1}{L^{D}}\iint d^{D}\mathbf{x}d^{D}\mathbf{y}  \notag \\
&&\times \hat{\Psi}_{1}^{\dagger }(\mathbf{x})\hat{\Psi}_{2}^{\dagger }(%
\mathbf{y})e^{i\mathbf{K\cdot }(\mathbf{x}+\mathbf{y})/2}|0\rangle .
\label{f}
\end{eqnarray}%
Here $\mathbf{K}$ is the total center-of-mass momentum, $L$ is the
quantization length in each dimension, and the states are normalized to one.
Calculating the matrix element in Eq. (\ref{Vma-def}) gives
\begin{equation}
|V_{ma}|^{2}=\frac{\hbar ^{2}\chi _{D}^{2}}{L^{D}}.  \label{Vma}
\end{equation}

Next, the density of two-atom states $D^{(2)}(\epsilon )$ at the total
dissociation energy $\epsilon =2\hbar |\Delta |$ is the same as the standard
density of single-atom states $D(E)$ at energy $E=\epsilon /2$ evaluated at
half the quantization volume $V=L^{3}$ for a single molecule (or area $A=L^{2}$ in 2D, or length $%
L $ in 1D); equivalently it is the same as half the density of single-atom
states evaluated for the same quantization volume, and is given by \cite%
{Dissociation-exp-Ketterle}
\begin{equation}
D^{(2)}(\epsilon )=\frac{1}{2}D(E)=\left\{
\begin{array}{ll}
\alpha L/(2\sqrt{E}),&\mathrm{(1D)}, \\
\\
\pi (\alpha L)^{2}/2,&\mathrm{(2D)}, \\
\\
\pi (\alpha L)^{3}\sqrt{E},&\mathrm{(3D)},%
\end{array}%
\right.  \label{densityofstates}
\end{equation}%
where $\alpha  =(2m_{1}/\hbar
^{2})^{1/2}/2\pi $ is a constant. Combining Eqs.~(\ref{Gamma-def}), (\ref{Vma}) and (\ref%
{densityofstates}), and expressing the final result in terms of the absolute
detuning $|\Delta |$, we obtain Eq.~(\ref{Gamma}) for the decay rate $\Gamma
$.

\section{Decay rate in the indistinguishable case}

\label{sect:appendix-B}

The initial $|i\rangle $ and final $|f\rangle $ states describing,
 a free molecule and a pair of free indistinguishable atoms respectively are
\begin{eqnarray}
|i\rangle &=&\frac{1}{L^{D/2}}\int d^{D}\mathbf{x}\hat{\Psi}_{0}^{\dagger }(%
\mathbf{x})e^{i\mathbf{K\cdot x}}|0\rangle , \\
|f\rangle &=&\frac{1}{\sqrt{2}\,L^{D}}\iint d^{D}\mathbf{x}d^{D}\mathbf{y}
\notag \\
&&\times \hat{\Psi}_{1}^{\dagger }(\mathbf{x})\hat{\Psi}_{1}^{\dagger }(%
\mathbf{y})e^{i\mathbf{K\cdot }(\mathbf{x}+\mathbf{y})/2}|0\rangle ,
\end{eqnarray}%
and are again normalized to one. Applying these to the transition
matrix element $V_{ma}=\langle f|\hat{H}_{int}|i\rangle $, where
$\hat{H}_{int}$ is the interaction term in the Hamiltonian
(\ref{Hindistinguish}), we obtain
\begin{equation}
|V_{ma}|^{2}=\frac{\hbar ^{2}\chi _{D}^{2}}{2L^{D}}.
\end{equation}
Accordingly, the molecular decay rate, $\Gamma^{\prime }=2\pi
|V_{ma}|^{2}D^{(2)}(\epsilon )/\hbar $, where $D^{(2)}(\epsilon )$
is given by Eq.~(\ref{densityofstates}), takes the form of
Eq.~(\ref{Gamma'}).

\section{Parametrization at large dissociation energy}

\label{sect:appendix-C}

Following the arguments of Ref.~\cite{Jack-Pu} we first take the
continuum limit of Eqs.~(\ref{PMFT}) and convert the sums over
$\mathbf{k}$ into integrals. Changing from the momentum variable to
the energy $E_{k}=\hbar ^{2}k^{2}/2m_{1}$ and then to $\hbar \Delta
_{k}=E_{k}+\hbar \Delta $ where we recall that $\Delta <0$, we can
rewrite the equation for $f(\tau )$ as
\begin{equation}
\frac{df(\tau )}{d\tau }=-\frac{\hbar }{\beta _{0}^{2}}\int_{-|\Delta
|}^{\infty }d\Delta _{k}\,D(\hbar \Delta _{k}-\hbar \Delta )m_{k}(\tau ),
\label{dfdtau2}
\end{equation}%
where $D(E)$ is the density of states Eq.~(\ref{densityofstates}).

At large $|\Delta |$ the resonance condition $\Delta _{k}=0$ (or $%
E_{k}=\hbar |\Delta |$) leads to the population of only those atomic
states that have absolute momenta $k=|\mathbf{k}|$ in a narrow range
around $k_{0}= \sqrt{2m_{1}|\Delta |/\hbar }$. The density of states
in this range of momenta is essentially constant and therefore
$D(E_{k})$  can be approximated by $D(\hbar |\Delta
|)$ and be taken out of the integral in Eq.~(\ref{dfdtau2}). At the
same time, the lower limit of the integral ($-|\Delta |$) can be
replaced by $-\infty $, which leads to the following result in terms
of dimensionless parameters
\begin{equation}
\frac{df(\tau )}{d\tau }\simeq -\frac{\hbar D(\hbar |\Delta |)}{\beta
_{0}^{2}t_{0}}\int_{-\infty }^{\infty }d\delta _{k}\,m(\delta _{k},\tau ),
\label{dfdtau3}
\end{equation}%
where $\delta _{k}=t_{0}\Delta _{k}/\hbar $ and $\,m(\delta _{k},\tau )$ is
a continuous function corresponding to $m_{k}(\tau )$ after making the
variable changes.

The coefficient in front of the integral in Eq.~(\ref{dfdtau3})
motivates the introduction of a dimensionless parameter $\Upsilon $
defined via $\hbar D(\hbar |\Delta |)/(\beta _{0}^{2}t_{0})\equiv
\sqrt{2/\Upsilon }$ and therefore
\begin{equation}
\Upsilon =\frac{2\beta _{0}^{4}t_{0}^{2}}{\hbar ^{2}[D(\hbar |\Delta |)]^{2}}%
=\frac{2N_{0}}{\hbar ^{2}\kappa ^{2}[D(\hbar |\Delta |)]^{2}}.
\end{equation}%
The final explicit result for $\Upsilon $ in 1D, 2D, and 3D is given
in Eq. (\ref{Gamma-Jack-Pu}).

\section{Width of the dissociation sphere}

\label{sect:appendix-D}

To estimate the width of the shell of the dissociation sphere beyond the spontaneous regime
we use the analytic solutions for a uniform system in the undepleted
molecular BEC approximation \cite{Fermidiss,Savage}. The solution for the
momentum distribution $n_{\mathbf{k}}(t)$ in the bosonic case reads as%
\begin{equation}
n_{\mathbf{k}}(t)=\frac{g^{2}}{g^{2}-\Delta _{k}^{2}}\sinh ^{2}\left( \sqrt{%
g^{2}-\Delta _{k}^{2}}\,t\right) ,  \label{n-k-analytic-b}
\end{equation}%
where%
\begin{eqnarray}
g &=&\chi _{D}\sqrt{N_{0}/L^{D}},  \label{g-def} \\
\Delta _{k} &\equiv &\frac{\hbar k^{2}}{2m}-\frac{\hbar k_{0}^{2}}{2m}.
\label{Delta-k-def}
\end{eqnarray}
From Eq.~(\ref{n-k-analytic-b}) we see that  modes with $g^{2}-\Delta _{k}^{2}>0$
experience Bose enhancement and grow exponentially with time, and the modes
with $g^{2}-\Delta _{k}^{2}<0$  oscillate at the spontaneous noise
level. The absolute momenta of the exponentially growing modes lie near the
resonant momentum $k_{0}$, and therefore we can use the condition $%
g^{2}-\Delta _{k}^{2}=0$ to define the width of the dissociation sphere.
First we write $k=k_{0}+\delta k$ and assume for simplicity that $k_{0}$ is
large enough so that $\delta k\ll k_{0}$. Then the above condition can be
approximated by
\begin{equation}
1-\left( \frac{\hbar k_{0}\delta k}{mg}\right) ^{2}\simeq 0.
\end{equation}%
This can be solved for $\delta k$ and used to define the width $\Delta
k=\delta k/2$ of the dissociation sphere as
\begin{equation}
\frac{\Delta k}{k_{0}}\simeq \frac{mg}{2\hbar k_{0}^{2}}=\frac{m\chi _{D}%
\sqrt{N_{0}/L^{D}}}{2\hbar k_{0}^{2}}.  \label{width}
\end{equation}%
The reason for defining it as half the full-width $\delta k$
is to make $\Delta k$ closer in definition to the rms width around
$k_{0}$.

For the case of fermionic statistics a similar analytic solution exists \cite%
{Fermidiss}%
\begin{equation}
n_{\mathbf{k}}(t)=\frac{g^{2}}{g^{2}+\Delta _{k}^{2}}\sin ^{2}\left( \sqrt{%
g^{2}+\Delta _{k}^{2}}\,t\right) ,
\end{equation}%
where $g$ and $\Delta _{k}$ are the same as in Eqs.~(\ref{g-def})
and (\ref{Delta-k-def}). In this case the solutions for all $k$ are
oscillatory, with populations of individual modes not
exceeding one due to the Pauli exclusion principle. The width of the
dissociation sphere can now be defined as the width of the envelope
function given by the Lorentzian $g^{2}/[g^{2}+\Delta _{k}^{2}]$.
Rewriting the width of this Lorentzian in terms of the momentum width
$\Delta k=\delta k/2$, we obtain the same expression as in
Eq.~(\ref{width}).


\begin{thebibliography}{99}
\bibitem{Yasuda-Shimizu} M. Yasuda and F. Shimizu, Phys. Rev. Lett. \textbf{%
77}, 3090 (1996).

\bibitem{burt97} E. A. Burt \textit{et al.}, Phys. Rev. Lett. \textbf{79},
337 (1997).

\bibitem{tolra04} B. Laburthe Tolra \textit{et al.}, Phys. Rev. Lett. \textbf{92},
190401 (2004).

\bibitem{kinoshita05} T. Kinoshita, T. Wenger, and D. S. Weiss, Phys. Rev.
Lett. \textbf{95}, 190406 (2005).

\bibitem{Greiner} M. Greiner \textit{et al.}, Phys. Rev. Lett. \textbf{94},
110401 (2005).

\bibitem{Bloch} S. F\"{o}lling \textit{et al.}, Nature (London)
\textbf{434}, 481 (2005).

\bibitem{Bloch-HBT-fermions} T. Rom \textit{et al.}, Nature (London),
\textbf{444}, 733 (2006).

\bibitem{esteve06} J. Esteve \textit{et al.}, Phys. Rev. Lett. \textbf{96},
130403 (2006).

\bibitem{Raizen} C.-S. Chuu \textit{et al.}, Phys. Rev. Lett. \textbf{95},
260403 (2005).

\bibitem{Esslinger} A. \"{O}ttl \textit{et al.}, Phys. Rev. Lett.
\textbf{95}, 090404 (2005).

\bibitem{Aspect} M. Schellekens \textit{et al.}, Science \textbf{310}, 648
(2005).

\bibitem{Aspect-HBT-fermions} T. Jeltes \textit{et al.}, Nature (London)
\textbf{445}, 402 (2007).

\bibitem{Perrin-BEC-collisions} A. Perrin \textit{et al.}, arXiv:0704.3047.

\bibitem{photoionization} T. Campey \textit{et al.}, Phys. Rev. A
\textbf{74}, 043612 (2006); S. Karft \textit{et al.}, A \textbf{75},
063605 (2007).

\bibitem{Sykes} A. G. Sykes, D. M. Gangardt, M. J. Davis, K. Viering,
M. G. Raizen, and K. V. Kheruntsyan, in preparation.

\bibitem{Trippenbach} R. Bach, M. Trippenbach, and K. Rz\c{a}\.{z}ewski, Phys.
Rev. A \textbf{65}, 063605 (2002); P. Zi\'{n} et al., Phys. Rev. Lett.
\textbf{94}, 200401 (2005).

\bibitem{Perrin-theory} A. Perrin, C. M. Savage, V. Krachmalnicoff, D.
Boiron, C. I. Westbrook, and K. V. Kheruntsyan, in preparation.

\bibitem{Moelmer2001} U. V. Poulsen and K. M\o lmer, Phys. Rev. A
\textbf{63}, 023604 (2001).

\bibitem{twinbeams} K. V. Kheruntsyan and P. D. Drummond, Phys. Rev. A
\textbf{66}, 031602(R) (2002); K. V. Kheruntsyan, Phys. Rev. A
\textbf{71}, 053609 (2005).

\bibitem{Vardi-Moore} A. Vardi and M. G. Moore, Phys. Rev. Lett.
\textbf{89}, 090403 (2002).

\bibitem{Norrie-Ballagh-Gardiner} A. A. Norrie, R. J. Ballagh, and C. W.
Gardiner, Phys. Rev. Lett. \textbf{94}, 040401 (2005); Phys. Rev. A \textbf{%
73}, 043617 (2006).

\bibitem{Rey-Clark} A. M. Rey, I. I. Satija, and C. W. Clark, J. Phys. B
\textbf{39}, S177 (2006); New J. Phys. \textbf{8}, 155 (2006).

\bibitem{Jack-Pu} M. W. Jack and H. Pu, Phys. Rev. A \textbf{72}, 063625
(2005).

\bibitem{Savage} C. M. Savage, P. E. Schwenn, and K. V. Kheruntsyan, Phys. Rev.
A \textbf{74}, 033620 (2006).

\bibitem{DeuarDrummond-4WM} P. Deuar and P. D. Drummond, Phys. Rev. Lett.
\textbf{98}, 120402 (2007).

\bibitem{Challis} K. J. Challis, R. J. Ballagh, and C. W. Gardiner,
Phys. Rev. Lett. \textbf{98}, 093002 (2007).

\bibitem{Kasevich} C. Orzel \textit{et al.}, Science 291, 2386 (2001).

\bibitem{Roberts} D. C. Roberts, T. Gasenzer and K. Burnett, J. Phys. B: At.
Mol. Opt. Phys. \textbf{35}, L113 (2002).

\bibitem{Yurovski-diss} V. A. Yurovsky and A. Ben-Reuven, Phys. Rev. A
\textbf{67}, 043611 (2003).

\bibitem{Haine} M. T. Johnsson and S. A. Haine, Phys. Rev. Lett.
\textbf{99}, 010401 (2007).

\bibitem{SavageKheruntsyanSpatial} C. M. Savage and K. V. Kheruntsyan,
arXiv:0705.4135v1 (Phys. Rev. Lett., in press).

\bibitem{Meystre-spin-EPR} H. Pu and P. Meystre, Phys. Rev. Lett.
\textbf{85}, 3987 (2000).

\bibitem{Duan-spin-EPR} L.-M. Duan \textit{et. al.}, Phys. Rev. Lett.
\textbf{85}, 3991 (2000).

\bibitem{Soerensen-Duan-Zoller} A. S\o rensen \textit{et. al.}, Nature
(London) \textbf{409}, 63 (2001).

\bibitem{Yurovski-FWM} V. A. Yurovsky, Phys. Rev. A \textbf{65}, 033605
(2002).

\bibitem{KPRL} K. V. Kheruntsyan, M. K. Olsen, and P. D. Drummond, Phys.
Rev. Lett. \textbf{95}, 150405 (2005).

\bibitem{Hope} S. A. Haine and J. J. Hope, Phys. Rev. A \textbf{72}, 033601
(2005).

\bibitem{Olsen-Davis} M. K. Olsen and M. J. Davis, Phys. Rev. A \textbf{73},
063618 (2006).

\bibitem{Zhao-Astrakharchik} B. Zhao \textit{et al.}, Phys. Rev. A
\textbf{75}, 042312 (2007).

\bibitem{Wieman-Julienne-dissociation} S. T. Thompson, E. Hodby, and C. E.
Wieman, Phys. Rev. Lett. \textbf{94}, 020401 (2005); T. K\"{o}hler,
E. Tiesinga, and P. S. Julienne, Phys. Rev. Lett. \textbf{94},
020402 (2005).

\bibitem{Rempe-Kokkelmans-dissociation} S. D\"{u}rr \textit{et al.}, Phys.
Rev. A \textbf{72}, 052707 (2005).

\bibitem{Braaten-2006} E. Braaten and D. Zhang, Phys. Rev. A \textbf{73},
042707 (2006).

\bibitem{Hanna-2006} T. M. Hanna, K. G\'{o}ral, E. Witkowska,
and T. K\"{o}hler, Phys. Rev. A \textbf{74}, 023618 (2006).

\bibitem{Vardi} I. Tikhonenkov and A. Vardi, Phys. Rev. Lett. \textbf{98},
080403 (2007).

\bibitem{Plata-2005} S. Brouard and J. Plata, Phys. Rev. A \textbf{72},
023620 (2005).

\bibitem{Meystre-diss} T. Miyakawa and P. Meystre, Phys. Rev. A \textbf{74},
043615 (2006).

\bibitem{Holland} M. Holland, J. Park, and R. Walser, Phys. Rev. Lett.
\textbf{86}, 1915 (2001); S. J. J. M. F. Kokkelmans and M. J. Holland, Phys.
Rev. Lett. \textbf{89}, 180401 (2002).

\bibitem{Gilchrist} A. Gilchrist, C. W. Gardiner, and P. D. Drummond,
Phys. Rev. A \textbf{55}, 3014 (1997); P. Deuar and P. D. Drummond,
J. Phys. A: Math. Gen. \textbf{39}, 1163 (2006).

\bibitem{Corney-fermionic} J. F. Corney and P. D. Drummond, Phys. Rev. Lett.
\textbf{93}, 260401 (2004); Phys. Rev. B \textbf{73}, 125112 (2006); P.
Corboz \textit{et al.}, arXiv:0707.4394.

\bibitem{Fermidiss} K. V. Kheruntsyan, Phys. Rev. Lett. \textbf{96}, 110401
(2006).

\bibitem{Dissociation-exp-Ketterle} T. Mukaiyama \textit{et al.}, Phys. Rev.
Lett. \textbf{92}, 180402 (2004).

\bibitem{Durr} S. D\"{u}rr, T. Volz, and G. Rempe, Phys. Rev. A \textbf{70},
031601(R) (2004).

\bibitem{JOptB1999-PRA2000} P. D. Drummond, K. V. Kheruntsyan, and H. He, J.
Opt. B: Quantum Semiclass. Opt. \textbf{1}, 387 (1999); K. V. Kheruntsyan
and P. D. Drummond, Phys. Rev. A \textbf{61}, 063816 (2000).

\bibitem{PDKKHH-1998} P. D. Drummond, K. V. Kheruntsyan, and H. He, Phys.
Rev. Lett. \textbf{81}, 3055 (1998); K. V. Kheruntsyan and P. D. Drummond,
Phys. Rev. A \textbf{58}, R2676 (1998); K. V. Kheruntsyan and P. D.
Drummond, Phys. Rev. A \textbf{58}, 2488 (1998).

\bibitem{Superchemistry} D. J. Heinzen, R. H. Wynar, P. D. Drummond, and K.
V. Kheruntsyan, Phys. Rev. Lett. \textbf{84}, 5029 (2000).

\bibitem{Timmermans} E. Timmermans \textit{et al.}, Phys. Rev. Lett. \textbf{%
83}, 2691 (1999); Phys. Rep. \textbf{315}, 199 (1999).

\bibitem{Feshbach-KKPD} P. D. Drummond and K. V. Kheruntsyan, Phys. Rev. A
\textbf{70}, 033609 (2004).

\bibitem{JJ-1999} J. Javanainen and M. Mackie, Phys. Rev. A \textbf{59},
R3186 (1999).

\bibitem{Stoof-review} R. A. Duine and H. T. C. Stoof, Phys. Rep.
\textbf{86}, 115 (2004).

\bibitem{Julienne-review} T. K\"{o}hler, K. G\'{o}ral, and P. S. Julienne,
Rev. Mod. Phys. \textbf{78}, 1311 (2006).

\bibitem{Mies} F. H. Mies \textit{et al.}, Phys. Rev. A \textbf{61}, 022721
(2000).

\bibitem{Comment1} Notationally, the equivalence is established by noting
that the detuning $\nu $ of Ref. \cite{Jack-Pu} corresponds to $-\Delta $
used here. In addition, we have found the following typographical errors in
Ref. \cite{Jack-Pu}: (\textit{i}) in Eq. (1), $\nu $ should be replaced by $%
2\nu $, and (\textit{ii}) in Figs. 2 and 3, the scale on the horizontal axis
should read as Time ($1/g\sqrt{N}$), which corresponds to $1/(\kappa \sqrt{%
2|\beta _{0}|^{2}})$ in our notation.

\bibitem{Glauber} R. J. Glauber, Phys. Rev. \textbf{130}, 2529 (1963);
M. Naraschewski and R. J. Glauber, Phys. Rev. A \textbf{59}, 4595
(1999).
\end{thebibliography}
\end{document}